\documentclass[letterpaper,11pt]{article}
\pdfoutput=1 

\usepackage{jheppub} 

\usepackage[english]{babel}
\usepackage[utf8]{inputenc}
\usepackage[T1]{fontenc} 
\usepackage[dvipsnames]{xcolor}
\usepackage{multirow}
\usepackage{slashed}
\usepackage[normalem]{ulem}
\usepackage{sidecap}
\usepackage{mathtools}
\usepackage{subcaption}
\usepackage[toc]{appendix}
\usepackage{amsmath}
\usepackage{ulem}
\usepackage{listings}
\usepackage{gensymb}
\usepackage{latexsym}
\usepackage{epsfig}
\usepackage{amsfonts,amssymb}
\usepackage{graphicx}
\usepackage{fontawesome5}
\usepackage{listings} 
\definecolor{keywordcolor}{RGB}{0,0,180}
\definecolor{classcolor}{RGB}{180,0,0}
\definecolor{functioncolor}{RGB}{0,128,0}
\definecolor{commentcolor}{RGB}{120,120,120}
\lstdefinelanguage{PythonCustom}{
  language=Python,
  morekeywords={self},                
  sensitive=true,
  morecomment=[l]{\#},
  morestring=[b]',
  morestring=[b]",
}
\usepackage{tikz}
\usepackage{tikz-feynman}
\tikzfeynmanset{compat=1.1.0}

\usepackage[backend=biber,style=nature,doi=true]{biblatex}
\title{Constraining light dark matter in vector-scalar portals with COSI and AMEGO-X}

\author[a,b]{Ma\'ira Dutra,}
\author[c]{Clarissa Siqueira,}
\author[a]{and Tonia Venters}

\affiliation[a]{Astroparticle Physics Laboratory, NASA Goddard Space Flight Center, Greenbelt, MD 20771, USA}
\affiliation[b]{NASA Postdoctoral Program Fellow}
\affiliation[c]{Observatório Nacional, 20921-400, Rio de Janeiro - RJ, Brazil}

\emailAdd{mdutrava@mail.nasa.gov}
\emailAdd{csiqueira@on.br}
\emailAdd{tonia.m.venters@nasa.gov}

\usepackage{calligra}
\usepackage{calrsfs}
\DeclareMathAlphabet{\mathcalligra}{T1}{calligra}{m}{n}
\DeclareMathAlphabet{\pazocal}{OMS}{zplm}{m}{n}

\def\dm{\text{\tiny{DM}}}

\def\zp{$Z'$ }

\def\Lag{\pazocal{L}}

\def\figureautorefname~#1\null{Fig.\,#1\null}
\def\tableautorefname~#1\null{Tab.\,#1\null}

\def\equationautorefname~#1\null{Eq.\,(#1)\null}

\abstract{Detecting gamma-ray signals that could be due to dark matter (DM) particles would give us invaluable information about the nature of DM. In particular, gamma-ray lines could provide a way to measure the DM mass. The excellent energy resolution of the upcoming Compton Spectrometer and Imager (COSI) will allow us to probe underexplored regions of the DM parameter space while being sensitive to distinctive spectral features of potential DM signals. In this work, we consider a fermionic sub-GeV DM charged under a new U(1) gauge symmetry. Both the DM and the new gauge boson $Z'$ acquire mass from a new singlet scalar. The masses of the new particles in this class of vector-scalar portal models are naturally at the MeV scale, enabling detectable gamma-ray lines in the bandpasses of COSI and proposed missions such as the All-sky Medium Energy Gamma-ray Observatory eXplorer (AMEGO-X). We estimate the sensitivities of COSI and AMEGO-X to sub-GeV DM in this context, considering a B-L and a purely axial $Z'$ as benchmark examples. We find regions of the parameter space where COSI will provide leading constraints, beyond the strong CMB limits. On the other hand, AMEGO-X would probe most of the viable parameter space leading to continuum gamma rays. The implementation of our generic vector-scalar portal model in the {\tt Hazma} toolkit is available at {\tt GitHub} \href{https://github.com/dutramaira/VectorScalarPortal_Hazma}{\faGithub}.}

\begin{document}

\maketitle
\flushbottom

\section{Introduction}
\label{sec:intro}

The presence of dark matter (DM) particles in dense regions of the universe today is well-established by astrophysical and cosmological observations \cite{Zwicky:1933gu,Harvey:2015hha,Aylor:2017haa,Aghanim:2018eyx,ACT:2020gnv} and implies new physics beyond the Standard Model (SM) of particle physics. However, we cannot know from those observations whether DM and SM particles interact non-gravitationally and what is the nature of DM, for instance its mass scale and quantum numbers. We have been searching for DM for more than forty years with no conclusive results \cite{Stecker:1978du,Goodman:1984dc,Sadoulet:2024geu,LZ:2022lsv,HESS:2022ygk,MAGIC:2016xys}, but some puzzling signals could be explained by non-gravitational interactions of DM \cite{Hooper:2010mq,Hooper:2011ti,2009APh....32..140G,Leane:2022bfm}. One of the most promising ways to search for DM is by looking for an excess of gamma rays coming from dense regions, such as the Galactic Center, relative to the emission expected by known astrophysical processes. In particular, DM annihilation or decay into photons, although usually suppressed relative to annihilation into leptons, would generate monochromatic gamma rays with energy equal to the DM mass. Such a signal is, in principle, much easier to distinguish from other known nuclear gamma-ray lines or the continuum gamma rays from diffuse emission, and would represent a major advance in the understanding of the nature of DM. 

The Compton Spectrometer and Imager (COSI) \cite{Tomsick:2023aue}, a NASA Small Explorer satellite scheduled for launch in 2027, will provide a unique opportunity in the search for DM. COSI is a wide field-of-view compact Compton telescope optimized for gamma-ray line searches in the 0.2-5 MeV range, with an excellent energy resolution of less than 1\%. Many other MeV gamma-ray telescopes are currently being developed and planned, such as the All-sky Medium Energy Gamma-ray Observatory eXplorer (AMEGO-X) \cite{Caputo:2022xpx} and the Galactic Explorer with a Coded Aperture Mask Compton telescope (GECCO) \cite{Orlando:2021get}. These telescopes aim to significantly improve the gap in sensitivity to MeV-scale gamma-ray emission, which is due to technological and theoretical challenges related to this transition energy range, as compared to X-ray and high-energy gamma-ray emissions. The consequence for DM phenomenology is that we will be able to explore new regions of the DM parameter space that are difficult to probe with other experiments \cite{Caputo:2017sjw,Bartels:2017dpb,Coogan:2021rez,Caputo:2022dkz,ODonnell:2024aaw,Linden:2024fby,Watanabe:2025pvc,Cirelli:2025qxx,Cirelli:2025rky,Fujisawa:2025yqi,Nguyen:2025tkl,Saha:2025wgg}.  

The vast majority of DM models and experimental searches concentrate on DM masses in the GeV-TeV scale. This is because neutral stable particles with masses and couplings at the electroweak scale are predicted by appealing extensions of the SM, with the relic abundance of such DM candidates being easily achieved via the well-understood freeze-out mechanism, and because we have a solid technology to search for these so-called weakly interacting massive particles (WIMPs). Even though the search for GeV-TeV WIMPs remains an important endeavor \cite{Arcadi:2024ukq}, the lack of new physics signals at the electroweak scale and the advent of new technologies motivate us to consider alternatives. In particular, the upcoming COSI telescope strongly motivate the study of sub-GeV DM candidates that can produce gamma-ray lines. 

In this work, we focus on a popular class of DM models in which a fermionic DM, $\chi$, is charged under a new local $U(1)_X$ symmetry. The interactions between DM and the SM particles are mediated by the gauge boson associated with $U(1)_X$, the $Z'$ boson. This is an important framework of new physics because $U(1)_X$ groups are part of the breaking pattern of many larger symmetries \cite{Cvetic:1997mbh, London:1986dk, Hewett:1988xc, Chun:2008by, Langacker:2008yv} invoked to solve different open questions in particle physics, and because the DM stability is guaranteed by symmetry. The annihilation of DM into photons ($\chi\bar\chi\to\gamma\gamma$), which leads to gamma-ray line signals, is only possible in this context when both the DM and the SM fermions have axial-vector couplings to the $Z'$ boson \cite{Duerr:2015wfa}. Interestingly, the self-consistency of this scenario strongly relates the masses of $\chi$ and $Z'$. In the simplest case, we need a singlet scalar field of similar mass in order to break the $U(1)_X$ symmetry and give mass to $Z'$, which also interacts with the chiral fermions \cite{Kahlhoefer:2015bea}. This scalar field ($h'$) mixes with the SM Higgs boson, acting as a second mediator for the DM-SM interactions. We thus study the gamma-ray signals of a sub-GeV fermionic DM in vector-scalar portal models with sub-GeV mediators $Z'$ and $h'$, focusing on DM masses within COSI's energy range, from hundreds of keV up to a few MeV.

The implementation of our generic vector-scalar portal model is available at \\ \href{https://github.com/dutramaira/VectorScalarPortal_Hazma}{https://github.com/dutramaira/VectorScalarPortal\_Hazma} \href{https://github.com/dutramaira/VectorScalarPortal_Hazma}{\faGithub}.

In Sec.~\ref{sec:model} we review the class of vector-scalar portal models and define the benchmark models considered in our analysis, namely $U(1)_{B-L}$ and $U(1)_A$. In Sec.~\ref{sec:gammarays} we discuss the different gamma-ray signals from light dark matter in the context of our benchmark models. General aspects of indirect detection and other constraints are discussed in Sections \ref{sec:ID} and \ref{sec:OtherConstraints}, respectively. We show and discuss our results in Sections \ref{sec:results} and \ref{sec:literature}, and conclude in Sec. \ref{sec:conclusions}. In Appendix \ref{sec:Hazma} we provide details regarding the implementation of our model in the numerical package {\tt Hazma} and highlight some differences between our vector-scalar portal and the pure vector and pure scalar portals shipped with {\tt Hazma}.

\section{Vector-scalar portal models}
\label{sec:model}

In this work, we are interested in a minimal but self-consistent model that predicts gamma-ray lines from dark matter annihilation in the MeV scale. Our dark matter candidate is a new Dirac fermion $\chi$ charged under a new gauge symmetry, $U(1)_X$. We assume this symmetry was spontaneously broken by the vacuum expectation value (vev) of a complex scalar $\Phi_s$ that also interacts with $\chi$. In this way, after the $U(1)_X$ breaking, both the DM and the new gauge boson $Z'$ will acquire masses, and the DM stability is protected by the remnant $Z_2$ symmetry.

The Lagrangian of the scalar sector reads
\begin{equation}
\Lag \supset |D_\mu \Phi|^2 + |D_\mu \Phi_s|^2 - V(\Phi,\Phi_s) \,,
\end{equation}
where $\Phi = (-i G^+, (\phi^0 + i G^0)/\sqrt{2})^T$ is the electroweak doublet and $\Phi_s = (\phi_s + i G_s)/\sqrt{2}$ is the extra scalar, singlet under the SM gauge group. Their kinetic terms are given by 
\begin{align}
& D_\mu \Phi = (\partial_\mu - i g \frac{\sigma^i}{2}W^i_\mu - i g' Y_\Phi B_\mu - i g_X X_\Phi B'_\mu ) \Phi, \\
& D_\mu \Phi_s = (\partial_\mu - i g_X X_{\Phi_s} B'_\mu ) \Phi_s \,,
\end{align}
with $g, g'$, and $g_X$ the gauge couplings of $SU(2)_L$, $U(1)_Y$, and $U(1)_X$, and $Y$ and $X$ the hypercharge and the $U(1)_X$ charge, respectively. The scalar potential is given by
\begin{equation}
V(\Phi, \Phi_s) = 
\mu^2 |\Phi|^2 + \lambda |\Phi|^4 + \mu_s^2 |\Phi_s|^2 + \lambda_s |\Phi_s|^4 + \lambda_{hs} |\Phi|^2 |\Phi_s|^2  \,.
\end{equation}

The singlet scalar develops a vev $v_s$ ($\phi_s \to \phi_s + v_s$), breaking $SU(3)_c \times SU(2)_L \times U(1)_Y \times U(1)_X$ to the SM gauge symmetry $SU(3)_c \times SU(2)_L \times U(1)_Y$, and the doublet scalar $\Phi$ develops a vev $v$ ($\phi^0 \to \phi^0 + v$), breaking the SM symmetry to $SU(3)_c \times U(1)_{em}$. The scalar potential then leads to a mixing between the neutral scalars:
\begin{equation}
V(\Phi, \Phi_s) \supset \frac{1}{2} 
\begin{pmatrix}
\phi^0 & \phi^0_s
\end{pmatrix}
\begin{pmatrix}
2\lambda v^2 & \lambda_{hs} v v_s \\
\lambda_{hs} v v_s & 2\lambda_s v_s^2
\end{pmatrix}
\begin{pmatrix}
\phi^0 \\ \phi^0_s
\end{pmatrix}
\,.
\end{equation}

We diagonalize this mass matrix with a rotation of angle $\alpha$, defined by $\tan(2\alpha)=\lambda_{hs}v v_s/(\lambda v^2-\lambda_s v_s^2)$, to find the physical scalars $h = c_\alpha \,\phi^0 + s_\alpha \,\phi^0_s$ and $h' = -s_\alpha \,\phi^0 + c_\alpha \,\phi^0_s$, where $s_\alpha$ and $c_\alpha$ are the sine and cosine of $\alpha$, with masses $m_h\simeq 125$ GeV and $m_{h'}$, respectively:
\begin{align}
& m_h^2 = \lambda v^2 +\lambda_s v_s^2 +\sqrt{(\lambda v^2-\lambda_s v_s^2)^2+ \lambda_{hs}^2v^2v_s^2}\,,\nonumber\\
& m_{h'}^2 = \lambda v^2 +\lambda_s v_s^2 -\sqrt{(\lambda v^2-\lambda_s v_s^2)^2+ \lambda_{hs}^2v^2v_s^2}\,. 
\end{align}

The quartic couplings are given in terms of the scalar masses and mixing angle:
\begin{align}
\lambda = \frac{c_\alpha^2 m_h^2 + s_\alpha^2 m_{h'}^2}{2v^2}, \, \lambda_s = \frac{s_\alpha^2 m_h^2 + c_\alpha^2 m_{h'}^2}{2v_s^2}, \, \lambda_{hs} = \frac{s_\alpha c_\alpha (m_h^2 - m_{h'}^2)}{v v_s} \,.
\label{Eq:lambdas}
\end{align}

After the electroweak symmetry breaking, we have in principle a mass mixing between the neutral massive gauge bosons:
\begin{equation}
|D_\mu \Phi|^2 + |D_\mu \Phi_S|^2 \supset \frac{1}{2} 
\begin{pmatrix}
Z_\mu^0 & B'_\mu
\end{pmatrix}
\begin{pmatrix}
m_{Z^0}^2 & -\Delta^2 \\
-\Delta^2 & m_{B'}^2
\end{pmatrix}
\begin{pmatrix}
Z_\mu^0 \\ B'_\mu
\end{pmatrix}
\,,
\end{equation}
where $m_{Z^0}^2=g_Z^2 v^2/4$, $m_{B'}^2=g_X^2(X_{\Phi_s}^2 v_s^2 + X_{\Phi}^2 v^2)$, and $\Delta^2=g_Z g_X X_\Phi v^2/2$, with $g_Z^2 = g^2+{g'}^2$. We diagonalize this mass matrix with a rotation of angle $\xi$, given by $\xi = \tan^{-1}\left(\frac{\Delta^2}{m_{Z'}^2 - m_{Z^0}^2}\right)$. 

The masses of the physical gauge bosons $Z = c_\xi Z^0 + s_\xi B'$, identified as the SM $Z$ boson, and $Z' = -s_\xi Z^0 + c_\xi B'$ are given respectively by 
\begin{align}
& m_Z^2 = \frac{1}{2}\left(m_{Z^0}^2+m_{B'}^2+\sqrt{(m_{Z^0}^2-m_{B'}^2)^2+4\Delta^2}\right)\,, \nonumber\\
& m_{Z'}^2 = \frac{1}{2}\left(m_{Z^0}^2+m_{B'}^2-\sqrt{(m_{Z^0}^2-m_{B'}^2)^2+4\Delta^2}\right)\,. 
\end{align}

Solving these equations for $m_Z^2$ and $m_{B'}^2$ and applying the definitions of $m_{Z^0}^2$, $m_{B'}^2$, and $\Delta^2$ gives us
\begin{align}
& m_Z^2 = \frac{m_{Z^0}^4-m_{Z^0}^2 m_{Z'}^2+\Delta^4}{m_{Z^0}^2-m_{B'}^2}\,, \nonumber\\
& v_s = \frac{m_{Z'}}{g_X X_{\Phi_s}}\sqrt{1+\frac{4X_\Phi^2 g_X^2 m_{Z^0}^2}{g_Z^2 (m_{Z^0}^2-m_{Z'}^2)}}\,. 
\label{eq:vs}
\end{align}

The fermionic sector contains the interactions of the SM fermions $f$ and the DM with $Z$ and $Z'$:
\begin{equation}
\begin{split}
\Lag &\supset \sum_f \left( \overline{f_L} i \slashed D f_L + \overline{f_R} i \slashed D f_R \right) + \overline{\chi_L} i \slashed D \chi_L + \overline{\chi_R} i \slashed D \chi_R + h.c. 
\\
& \supset Z_\mu \bar f \gamma^\mu [(c_\xi \frac{g_Z}{2}V^{EW}_f-s_\xi g_X V_f^X)-\gamma_5(c_\xi \frac{g_Z}{2}A^{EW}_f-s_\xi g_X A_f^X) ] f\\
&~~~ +Z'_\mu \bar f \gamma^\mu [(s_\xi \frac{g_Z}{2}V^{EW}_f+c_\xi g_X V_f^X)-\gamma_5(s_\xi \frac{g_Z}{2}A^{EW}_f+c_\xi g_X A_f^X) ] f \\ &~~~ -s_\xi g_X Z_\mu \bar \chi \gamma^\mu (V_\chi-\gamma_5 A_\chi ) \chi + c_\xi g_X Z'_\mu \bar \chi \gamma^\mu (V_\chi-\gamma_5 A_\chi ) \chi \,,
\end{split} \label{eq:Lfermions}
\end{equation}
with vector couplings defined as $V_f^{EW} = t_{3_f} - 2 s_W^2 Q_f$, $V_f^X = (X_{f_L}+X_{f_R})/2$, and $V_\chi = (X_{\chi_L}+X_{\chi_R})/2$, and axial-vector couplings defined as $A_f^{EW} = t_{3_f}$, $A_f^X = (X_{f_L}-X_{f_R})/2$, and $A_\chi = (X_{\chi_L}-X_{\chi_R})/2$. In these definitions, $t_{3_f}$ and $Q_f$ are the eigenvalue of the third generator of $SU(2)_L$ and the electric charge for the SM fermion $f$, and $s_W$ is the sine of the weak mixing angle. For later convenience, we also define the vector and vector-axial coupling strengths of SM fermions with the $Z$ boson as $V_{fZ} = c_\xi \frac{g_Z}{2}V^{EW}_f-s_\xi g_X V_f^X$ and $A_{fZ} = c_\xi \frac{g_Z}{2}A^{EW}_f-s_\xi g_X A_f^X$, and with the $Z'$ boson as $V_{fZ'} = s_\xi \frac{g_Z}{2}V^{EW}_f+c_\xi g_X V_f^X$ and $A_{fZ'} = s_\xi \frac{g_Z}{2}A^{EW}_f+c_\xi g_X A_f^X$.

Finally, we assume that $\Phi_s$ gives mass to DM through a Yukawa coupling. Thus, the Yukawa sector containing the interactions among fermions and scalars reads 
\begin{align}
\Lag &\supset - \sum_{i,j=1}^3 ( Y^e_{ij} \overline{{L_L}_i} \Phi {l_R}_j + Y^d_{ij} \overline{ {Q_L}_i} \Phi {d_R}_j + Y^u_{ij} \overline{{Q_L}_i} \tilde \Phi {u_R}_j + y_\chi \Phi_s \overline{\chi_L} \chi_R + h.c. ) \nonumber \\
& \supset \sum_{f} \left( - c_\alpha \frac{m_f}{v} \bar f f h + s_\alpha \frac{m_f}{v} \bar f f h'\right) - s_\alpha \frac{m_\chi}{v_s} \bar \chi \chi h - c_\alpha \frac{m_\chi}{v_s} \bar \chi \chi h'.
\label{Eq:Yuk}
\end{align}

In this way, after the $U(1)_X$ breaking, DM will acquire mass given by $m_\chi = y_\chi v_s/\sqrt{2}$ and its stability will be protected by the remnant $Z_2$ symmetry provided that $X_{\chi_R} \neq \pm (X_{L_L}+X_{\Phi})$, so to avoid a Yukawa term among $\chi_R$, $\Phi$, and $L_L$. Gauge invariance of the Yukawa sector leads to additional constraints on the fermion charges: $X_{d_R}=X_{Q_L} - X_\Phi$, $X_{u_R}=X_{Q_L} + X_\Phi$, and $X_{l_R}=X_{L_L} - X_\Phi$. It is worth mentioning that DM might also be vector-like under $U(1)_X$, so that its interaction with $\Phi_s$ is not possible. In this case, DM could acquire mass from the interaction with an extra scalar field neutral under $U(1)_X$. As we will show in Sec.~\ref{sec:results}, this would have important phenomenological consequences. 

In this work, we consider two benchmark models leading to different phenomenological results: $U(1)_{B-L}$ and $U(1)_A$, with charge assignments summarized in Table \ref{tab:charges}. 

\begin{table}[]
\centering
\begin{tabular}{c|ccccccccc|}
\cline{2-10}
 &
  \multicolumn{1}{c|}{$L_L$} &
  \multicolumn{1}{c|}{$l_R$} &
  \multicolumn{1}{c|}{$Q_L$} &
  \multicolumn{1}{c|}{$u_R$} &
  \multicolumn{1}{c|}{$d_R$} &
  \multicolumn{1}{c|}{$\Phi$} &
  \multicolumn{1}{c|}{$\Phi_s$} &
  \multicolumn{1}{c|}{$\chi_L$} &
  \multicolumn{1}{c|}{$\chi_R$} 
  \\ \hline
\multicolumn{1}{|c|}{$SU(3)_c$}    & 1    & 1  & 3   & 1   & 1    & 1   & 1 & 1 & 1 
\\ \cline{1-1}
\multicolumn{1}{|c|}{$SU(2)_L$}    & 2    & 1  & 2   & 1   & 1    & 2   & 1 & 1 & 1 
\\ \cline{1-1}
\multicolumn{1}{|c|}{$U(1)_Y$}     & -1 & -2 & 1/3 & 4/3 & -2/3 & 1 & 0 & 0 & 0 
\\ \cline{1-1}
\hline
\multicolumn{1}{|c|}{$U(1)_{B-L}$} & -1   & -1 & 1/3 & 1/3 & 1/3  & 0   & 2 & 4 & 2 
\\ \cline{1-1}
\multicolumn{1}{|c|}{$U(1)_{A}$} & 1/3   & -1/3 & 1/3 & -1/3 & -1/3  & 2/3   & 2 & 4 & 2 
\\ \hline
\end{tabular}
\caption{Representations and charge assignment of the relevant fields under the gauge group $SU(3)_c \times SU(2)_L \times U(1)_Y \times U(1)_X$ with $U(1)_X = U(1)_{B-L}$ and $U(1)_X = U(1)_{A}$.}
\label{tab:charges}
\end{table}

The $U(1)_{B-L}$ model is one of the simplest anomaly-free setups for a new gauge symmetry. The gauge anomalies can be canceled by the addition of three right-handed neutrinos $N_R^i$ \cite{Carena:2004xs}, which can generate the active neutrino masses and mixing. In a minimal scenario, $\Phi_s$ gives Majorana masses to $N_R^i$ and to the SM neutrinos $\nu_\alpha$ via the type-I seesaw mechanism, which implies that the singlet scalar must carry lepton number $L=-2$ \cite{Duerr:2015wfa,Escudero:2018fwn}, and thus $B-L$ charge $X_{\Phi_s}=2$. On the other hand, since only one scalar doublet is present, it must be neutral under the new symmetry ($X_\Phi = 0$) to properly give mass to the SM fermions in a gauge invariant way \cite{Carena:2004xs}. As a consequence, there is no mass mixing between the gauge bosons ($\tan \xi \propto X_\Phi$) and the \zp mass is simply given by $m_{Z'}=X_{\Phi_s} g_X v_s$. Note that in this case, the SM fermions have only vector couplings ($X_{f_L} = X_{f_R}$). Since we must have $X_{\chi_R} \neq \pm 1$, we set $X_{\chi_L}=4$ such that $X_{\chi_R}= X_{\chi_L}-X_{\Phi_s} = 2$.

Models with more than one doublet scalar and a larger set of chiral fermions lead to the possibility of axial-vector couplings between the SM fermions and $Z'$. In particular, a $U(1)_A$ model was studied in Ref.\cite{Hooper:2014fda} to explain the Galactic Center GeV excess from p-wave annihilation of DM particles in a $Z'$ portal. Moreover, the DM annihilation via purely axial $Z'$ is known to evade the strong direct detection limits as the scattering off nuclei is spin-dependent. Therefore, we consider the SM fermions' charge assignment as in the $U(1)_A$ model. As we will show, such axial-vector couplings enable detectable gamma-ray lines from our DM candidate at the MeV scale.

In both cases, the relevant free parameters are $g_X, X_\Phi, X_{\Phi_s}, s_\alpha, m_{h'}$ and $m_{Z'}$ from the scalar sector, $X_{Q_L}$, $X_{L_L}$, and $X_{\chi_L}$ from the fermionic sector \footnote{For simplicity, we have assumed the same X charges for quarks and the same X charges for leptons, but a further generalization is straightforward.}, and $m_\chi$ from the Yukawa sector. For simplicity, we assume that the fields required by anomaly cancellation, such as right-handed neutrinos, are too heavy to be relevant for the DM phenomenology. The remaining free parameters are fixed by their SM values.  

\begin{figure}[t!]
\centering
\begin{subfigure}{0.2\textwidth}
\centering
\begin{tikzpicture}[scale=0.5]
 \begin{feynman}
    \vertex (i1) at (-2,1) {\(\chi\)};
    \vertex (i2) at (-2,-1) {\(\bar{\chi}\)};
    \vertex (a) at (0,0);
    \vertex (b) at (2,0); 
    \vertex (f1) at (4,1) {\(\gamma\)};
    \vertex (f2) at (4,-1) {\(\gamma\)};
    \diagram* {
      (i1) -- [fermion] (a) -- [fermion] (i2),
      (a) -- [draw=gray, line width=1.5pt, edge label={\(Z', h'\)}] (b), 
      (b) -- [photon] (f1),
      (b) -- [photon] (f2),
    };
  \end{feynman}
  \filldraw (b) circle (4pt);
\end{tikzpicture}
\caption{}
\end{subfigure}
\hspace{1.5cm}
\begin{subfigure}{0.2\textwidth}
\centering
\begin{tikzpicture}[scale=0.5]
  \begin{feynman}
    \vertex (i1) at (-2,1) {\(\chi\)};
    \vertex (i2) at (-2,-1) {\(\bar{\chi}\)};
    \vertex (a) at (0,0);
    \vertex (b) at (2,0);
    \vertex (f1) at (4.5,1) {\(h', \pi^0\)};
    \vertex (f2) at (4,-1) {\(\gamma\)};
    \diagram* {
      (i1) -- [fermion] (a) -- [fermion] (i2),
      (a) -- [boson, edge label={\(Z'\)}] (b),
      (b) -- [scalar] (f1),
      (b) -- [photon] (f2),
    };
  \end{feynman}
  \filldraw (b) circle (4pt);
\end{tikzpicture}
\caption{}
\end{subfigure}
\hspace{1.5cm}
\begin{subfigure}{0.2\textwidth}
\centering
\begin{tikzpicture}[scale=0.5]
  \begin{feynman}
    \vertex (i1) at (-2,1) {\(\chi\)};
    \vertex (i2) at (-2,-1) {\(\bar{\chi}\)};
    \vertex (a) at (0,0);
    \vertex (b) at (2,0);
    \vertex (f1) at (4,1) {\(Z'\)};
    \vertex (f2) at (4,-1) {\(\gamma\)};
    \diagram* {
      (i1) -- [fermion] (a) -- [fermion] (i2),
      (a) -- [scalar, edge label={\(h'\)}] (b),
      (b) -- [boson] (f1),
      (b) -- [photon] (f2),
    };
  \end{feynman}
  \filldraw (b) circle (4pt);
\end{tikzpicture}
\caption{}
\end{subfigure}

\vspace{0.2cm}

\begin{subfigure}{0.2\textwidth}
\centering
\begin{tikzpicture}[scale=0.5]
  \begin{feynman}
    \vertex (i1) at (-2,1) {\(\chi\)};
    \vertex (i2) at (-2,-1) {\(\chi\)};
    \vertex (a) at (0.5,1);
    \vertex (b) at (0.5,-1);
    \vertex (c) at (2,1);
    \vertex (d) at (2,-1);
    \vertex (f1) at (4,1) {\(h',Z',h'\)};
    \vertex (f2) at (4,-1) {\(h',Z',Z'\)};
    
    \diagram* {
      (i1) -- [fermion] (a) -- [fermion, edge label=\(\chi\)] (b) -- [fermion] (i2),
      (a) -- [draw=gray, line width=1.5pt] (c) -- [draw=gray, line width=1.5pt] (f1),
      (b) -- [draw=gray, line width=1.5pt] (d) -- [draw=gray, line width=1.5pt] (f2),
    };
  \end{feynman}
\end{tikzpicture}
\caption{}
\end{subfigure}
\hspace{1.5cm}
\begin{subfigure}{0.2\textwidth}
\centering
\begin{tikzpicture}[scale=0.5]
  \begin{feynman}
    \vertex (i1) at (-2,1) {\(\chi\)};
    \vertex (i2) at (-2,-1) {\(\bar{\chi}\)};
    \vertex (a) at (0,0);
    \vertex (b) at (2,0);
    \vertex (f1) at (4.5,1) {\(h',Z'\)};
    \vertex (f2) at (4.5,-1) {\(h',Z'\)};
    \diagram* {
      (i1) -- [fermion] (a) -- [fermion] (i2),
      (a) -- [scalar, edge label={\(h'\)}] (b),
      (b) -- [draw=gray, line width=1.5pt] (f1),
      (b) -- [draw=gray, line width=1.5pt] (f2),
    };
  \end{feynman}
\end{tikzpicture}
\caption{}
\end{subfigure}
\hspace{1.5cm}
\begin{subfigure}{0.2\textwidth}
\centering
\begin{tikzpicture}[scale=0.5]
  \begin{feynman}
    \vertex (i1) at (-2,1) {\(\chi\)};
    \vertex (i2) at (-2,-1) {\(\bar{\chi}\)};
    \vertex (a) at (0,0);
    \vertex (b) at (2,0);
    \vertex (f1) at (4,1) {\(h'\)};
    \vertex (f2) at (4,-1) {\(Z'\)};
    \diagram* {
      (i1) -- [fermion] (a) -- [fermion] (i2),
      (a) -- [boson, edge label={\(Z'\)}] (b),
      (b) -- [scalar] (f1),
      (b) -- [boson] (f2),
    };
  \end{feynman}
\end{tikzpicture}
\caption{}
\end{subfigure}

\vspace{0.2cm}

\begin{subfigure}{0.45\textwidth}
\centering
\begin{tikzpicture}[scale=0.5]
  \begin{feynman}
    \vertex (i1) at (-2,1) {\(\chi\)};
    \vertex (i2) at (-2,-1) {\(\bar{\chi}\)};
    \vertex (a) at (0,0);
    \vertex (b) at (2,0);
    \vertex (f1) at (4,1) {\(l^-\)};
    \vertex (f2) at (4,-1) {\(l^+\)};
    \vertex (v1) at (3,0.5); 
    \vertex (g1) at (5,0.5) {\(\gamma\)}; 
    \diagram* {
      (i1) -- [fermion] (a) -- [fermion] (i2),
      (a) -- [draw=gray, line width=1.5pt, edge label={\(Z', h'\)}] (b), 
      (f2) -- [fermion] (b) -- [fermion] (f1),
      (v1) -- [photon] (g1), 
    };
  \end{feynman}
\end{tikzpicture}
\caption{}
\end{subfigure}

\caption{Relevant processes leading to gamma-ray signals in vector-scalar portal models. The black dots represent effective interactions and the gray tick lines represent a vector or scalar particle. For simplicity, we omit the exchanges of the SM $h$ and $Z$, since their contributions to the cross-sections are subdominant.}
\label{fig:diagrams}
\end{figure}
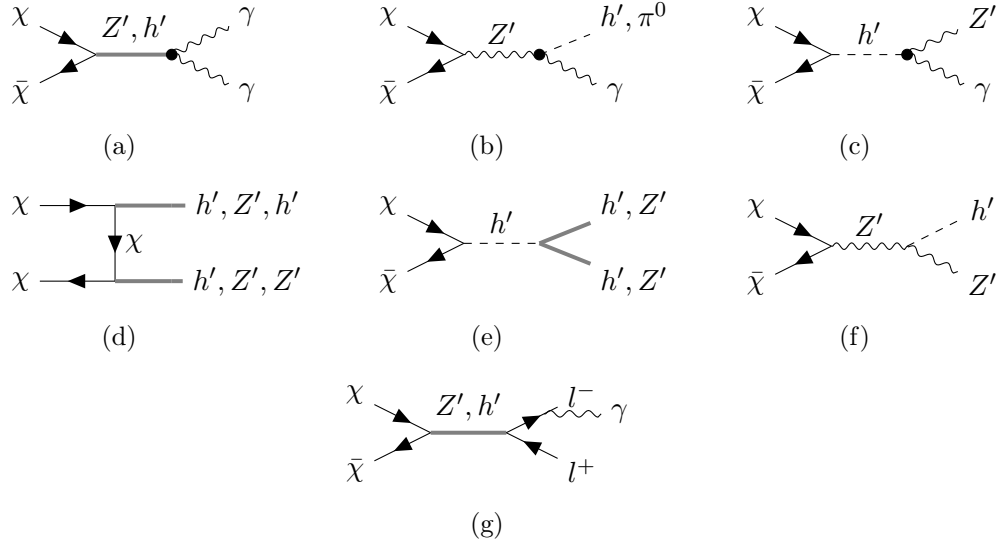

The DM annihilation channels leading to gamma rays in vector-scalar portal models are shown in Fig.\,\ref{fig:diagrams}. DM can annihilate directly into photons (via loop), panels (a)-(c), into dark sector particles, panels (d)-(f), through $t$ and $s$ channels, and into SM fermions, panel (g). 

We have implemented our model in the {\tt FeynRules} software \cite{Alloul:2013bka} to compute all the tree-level vertices needed for our numerical analyses. Let us now discuss the loop-induced vertices of panels (a)-(c). We provide below general expressions and the most relevant references where interested readers can find further details.

The effective coupling between a CP-even scalar $S$, a vector boson $V$, and a photon $\gamma$ is induced by loops of $W$ bosons and fermions $f$ and is given by \cite{Hue:2017cph}
\begin{equation}
(S, V_\mu (p_1), \gamma_\nu (p_2))^{\mu\nu}= (F^W_{SV} + F^f_{SV})[-(p_1\cdot p_2) g^{\mu\nu}+{p_2}^\mu{p_1}^\nu]\,,
\end{equation}
where the factor for the $W$ contribution is 
\begin{equation}
\begin{split}
F^W_{SV} = \frac{e}{8\pi^2} g_{SW} g_{VW} \Big\{ & C_0(m_S,m_V,m_W) \left(8-\frac{2m_V^2}{m_W^2}\right) \\
& + C_1 (m_S,m_V,m_W) \left[ 12 - \frac{2m_V^2}{m_W^2} + \frac{m_S^2}{m_W^2}\left(2-\frac{m_V^2}{m_W^2}\right) \right]  \Big\}
\end{split}
\end{equation}
and the factor for the $f$ contribution is 
\begin{equation}
\begin{split}
F^f_{SV} = -\frac{e}{2\pi^2} N_c Q_f m_f Y_{Sf} V_f^X \Big[ C_0(m_S,m_V,m_f) + 4 C_1(m_S,m_V,m_f) \Big] \,,
\end{split}
\end{equation}
with $C_0(m_a,m_b,m_c)=-I_2(t_a,t_b)/m_c^2$ and $C_1(m_a,m_b,m_c)=I_1(t_a,t_b)/(4m_c^2)$ the widely-used loop functions defined in Appendix A of Ref.\,\cite{Hue:2017cph}. 

The expressions above are given in terms of tree-level couplings $g_{SW}$ (for $SWW$), $g_{VW}$ (for $VWW$), and $Y_{Sf}$ (for $S\bar ff$), as well as the number of colors $N_c$ and the electric charge $Q_f$ of the fermions in the loop. Note that only the vector couplings $V_f^X$ contribute to the amplitudes. It is straightforward to use this result to compute the effective couplings $h'\gamma\gamma$ and $Z'h'\gamma$ (as well as $h\gamma\gamma$ and $Z h'\gamma$) in Fig.\,\ref{fig:diagrams}. 

Charged fermions that have axial-vector couplings to $Z'$ ($A_f^X \neq 0$) induce the effective coupling $Z'\gamma\gamma$. In the case of a SM fermion $f$, we have \cite{Duerr:2015wfa}
\begin{align}
(Z'_\sigma, \gamma_\mu (p_1), \gamma_\nu (p_2) )^{\sigma\mu\nu}_f =  i & \frac{e^2 Q_f^2 g_X A_f^X N_c}{4\pi^2} \nonumber \\
& \times \Big\{ F_1^f \Big[ \epsilon^{\mu\nu\sigma\alpha} (p_1-p_2)_\alpha - \frac{2}{s}\epsilon^{\mu\sigma\alpha\beta} {p_1}_\alpha {p_2}_\beta p_1^\nu 
+ \frac{2}{s}\epsilon^{\nu\sigma\alpha\beta} {p_1}_\alpha {p_2}_\beta p_1^\mu \Big] \nonumber \\
& \hspace{.8cm}  + F_2^f \frac{2}{s} \epsilon^{\mu\nu\alpha\beta} {p_1}_\alpha {p_2}_\beta (p_1 + p_2)^\sigma   \Big\}\,,
\end{align}
with
\begin{equation}
\begin{split}
& F_1^f = 3 + \Lambda(s,m_f) + 2 m_f^2 C_0 (s,m_f)  \\
& F_2^f = 2 + \Lambda(s,m_f) 
\end{split}
\end{equation}
and the loop functions 
\begin{equation}
\begin{split}
& \Lambda(s,m_f) = \sqrt{1-4m_f^2/s} \ln{\frac{2m_f^2}{2m_f^2 - s (1+\sqrt{1-4m_f^2/s})}}   \\
& C_0 (s,m_f) = \frac{1}{2s} \ln^2\left( \frac{\sqrt{1-4m_f^2/s}-1}{\sqrt{1-4m_f^2/s}+1} \right)
\end{split}
\end{equation}

In models with charged beyond the SM (BSM) fermions axially coupled to $Z'$, there is an additional contribution to the coupling $Z'\gamma\gamma$ \cite{Dudas:2013sia}:
\begin{align}
(Z'_\sigma(p_{Z'}), \gamma_\mu (p_1), \gamma_\nu (p_2) )^{\sigma\mu\nu}_\Lambda 
& = \frac{g_X}{2\Lambda^2}\partial^\sigma Z'_\sigma \tilde F^{\mu\nu}F_{\mu\nu} \nonumber \\
& = -\frac{g_X}{\Lambda^2}p_{Z'}^\sigma \epsilon^{\mu\nu\alpha\beta} {p_1}_\alpha {p_2}_\beta \,,
\end{align}
where $\Lambda$ is the UV cutoff scale, $F_{\mu\nu}$ is the photon field strength tensor, and $\tilde F^{\mu\nu} = \frac{1}{2}\epsilon^{\mu \nu \alpha \beta} F_{\alpha\beta}$ is the dual field strength tensor. 

The effective coupling between a vector boson $V$, a neutral pion $\pi^0$, and a photon depends only on the vector couplings of the quarks running in the loops and can be parameterized as \cite{ARNELLOS1982378,Fujiwara:1985,Ko:1995it,Tulin:2014tya,Coogan:2019qpu}:
\begin{equation}
(V_\mu(p_1), \pi^0, \gamma_\nu (p_2))^{\mu\nu}= -i \frac{e}{2\pi^2 f_\pi} V_{qV} \epsilon^{\mu\nu\alpha\beta} {p_1}_\alpha {p_2}_\beta F_w(s) \,,
\end{equation}
where $F_w(s)=1/(1-s/m_w^2-i\Gamma_w/m_w)$ is the form factor from the meson $w$ propagator \cite{Tulin:2014tya}. The interaction strengths between the quarks and the vector bosons $Z$ and $Z'$, $V_{qV}$, are defined from Eq.\,\ref{eq:Lfermions}.

\section{Gamma rays from sub-GeV dark matter annihilation}
\label{sec:gammarays}

The differential gamma-ray flux from DM annihilation coming from a region of observation subtended by the solid angle $\Delta \Omega$ is given by
\begin{equation}\label{eq:fluxDM}
    \frac{d\Phi}{dE_\gamma} (E_\gamma, \Delta \Omega) \Big|_{\bar \chi \chi} = \frac{\Delta \Omega}{4\pi} \bar J (\Delta \Omega)  \frac{\langle \sigma v \rangle}{2 f_\chi m_\chi^2} \frac{dN}{dE_\gamma} (E_\gamma)\,, 
\end{equation}
where $E_\gamma$ is the photon energy; the observation region is $\Delta \Omega = 2\pi (1-\cos \theta_\text{obs})$ for a disk of angle $\theta_\text{obs}$ and $\Delta \Omega = (l_\text{max} - l_\text{min})(\sin b_\text{max} - \sin b_\text{min})$ for an $l \times b$ region; the averaged J-factor $\bar J(\Delta \Omega)$ quantifies the amount of DM pairs within the angular size $\Delta \Omega$; $\langle \sigma v \rangle$ is the total velocity-averaged DM annihilation cross-section; $f_\chi = 1$ ($f_\chi = 2$) if DM is (not) self-conjugate; and $dN/dE_\gamma$ is the total spectrum of gamma rays from DM annihilation.     

The averaged J-factor is defined as
\begin{align}
   \bar J(\Delta \Omega) &\equiv \frac{1}{\Delta \Omega}\int_{\Delta \Omega} d\Omega \int_{l.o.s.} ds \rho_\dm^2(r(s,\theta)) 
\end{align}
with $\rho_\dm(r)$ the DM density profile. The radial coordinate connects the Galactic Center (GC) to the target, $r(s,\theta)=\sqrt{r_\odot^2+s^2-2r_\odot s \cos \theta}$, with $\theta$ the angle between the line-of-sight $s$ (Sun-target) and the axis $\mathbf{r_\odot}$ (Sun-GC). For an $l\times b$ region, $\cos \theta = \cos{b}\cos{l}$. In this work, we will consider the Einasto DM profile as a benchmark and adopt the corresponding halo parameters derived in Ref.~\cite{de_Salas_2019}. 

For non-relativistic annihilation given as a sum of s-wave (independent of DM velocity) and p-wave (proportional to the squared DM velocity) contributions, the velocity-averaged cross-section is approximately $\langle \sigma v \rangle \approx \sigma v_{rel}$, with the DM pair relative velocity $v_{rel} = \sqrt{3} v_0$ and $v_0 \simeq 220$km/s the most probable velocity in the Milky Way \cite{Ambrogi:2018jqj}. The cross-sections are computed at the center-of-mass energy $E_{cm} = \sqrt{s} \simeq 2m_\chi (1+v_{rel}^2/8)$.

The total spectrum is the sum of the spectra from all kinematically available annihilation final states $f$. A gamma ray with energy $E$ is observed with an energy $E_\gamma$ due to the detector's finite energy resolution $\epsilon (E) = \Delta E/E$. The convolved total spectrum is thus given by 
\begin{equation}
    \frac{dN}{dE_\gamma}(E_\gamma) = \sum_f BR(\chi \bar\chi \to f) \int dE R_\epsilon (E_\gamma|E) \frac{dN^f}{dE} (E) \,,
\end{equation}
where $BR(\chi \bar\chi \to f)$ and $\frac{dN^f}{dE}(E)$ are respectively the annihilation branching fraction and the spectrum for the final state $f$ and $R_\epsilon (E_\gamma|E)$ is the spectral resolution function, approximated as a Gaussian of standard deviation $\sigma = \epsilon(E) E$\footnote{The energy resolution is sometimes given in terms of the full width at half maximum (FWHM), such that $\epsilon(E) = \frac{\sigma}{E} = \frac{1}{2\sqrt{2\ln 2}}\frac{FWHM}{E}$.}:
\begin{equation}
    R_\epsilon (E_\gamma|E) = \frac{1}{\sqrt{2\pi} \epsilon(E) E} \exp{\left[-\frac{(E_\gamma-E)^2}{2(\epsilon(E) E)^2}\right]} \,.
\end{equation}

In any given model, the annihilation of dark matter generates gamma rays with different spectral features:
\begin{itemize}
\item Gamma-ray lines: the direct annihilation into photons, $\chi \bar \chi \to Y \gamma$, leads to monochromatic gamma rays at energy $E \simeq m_\chi \left(1 - \frac{m_Y^2}{4m_\chi^2} \right)$;
\item Box-shaped spectra: the cascade annihilation into photons from the decay of a final state particle, $\chi \bar \chi \to A B$ followed by $A$ and/or B decaying into $\gamma \gamma$, leads to a box-shaped spectrum of width $\Delta E \simeq \sqrt{m_\chi^2-m_{A,B}^2}$; 
\item Continuum spectra from FSR: the direct annihilation into charged particles, $\chi \bar \chi \to l^- l^+$, leads to final state radiation (FSR) with a cut-off at $E_\gamma \sim m_\chi$;
\item Continuum spectra from cascade FSR: the cascade annihilation into neutral particles, $\chi \bar \chi \to A B$ followed by $A$ and/or $B$ decaying into charged particles, leads to a continuum spectrum with a cut-off at $m_\chi$ similar to the direct FSR case but softer. This contribution becomes relevant when DM interacts more strongly with the intermediate particles A and B than with charged SM leptons.  
\end{itemize}
 
The gamma-ray line spectra are simply given by 
\begin{equation}
    \frac{dN^{Y\gamma}}{dE}(E) = N_\gamma  \delta\left(\frac{E_{cm}^2-m_Y^2}{2E_{cm}}-E\right)\,,
\end{equation}
with $N_\gamma$ the number of photons produced per annihilation.

The box-shaped spectra are found after performing a boost of the decay spectra for $A \to \gamma\gamma$, given by $2 BR(A\to\gamma\gamma) \delta\left(m_A/2 - E_R \right)$, from the rest frame of the decaying particle $A$ to the DM frame, assumed to be comoving to the Galactic frame. $E_R = \gamma E (1-\beta \cos \theta_R)$ is the photon energy in the rest frame of $A$, $\theta_R$ is the angle between the photon direction in the $A$ frame and the boost direction, $\gamma = E_A/m_A$ is the boost factor, $E_A = (E_{cm}^2+m_A^2-m_B^2)/(2E_{cm})$ is the energy of the decaying particle in the DM frame, $m_{A,B}$ are the masses of the annihilation final states ($\chi\bar\chi \to A B$), and $\beta = \sqrt{1-1/\gamma^2}$ is the boost velocity. The resulting spectrum due to the process $\chi\bar\chi\to AB$ followed by $A\to\gamma\gamma$ is given by \cite{Coogan:2019qpu}
\begin{equation}
    \frac{dN^{AB}}{dE}(E)\Big|^{\gamma\gamma} = BR(A\to\gamma\gamma) \frac{2}{\gamma \beta m_A} \left[ \Theta(E-E_-) + \Theta(E_+-E) \right]\,,
\label{eq:CascadeDecay}
\end{equation}
where $E_\pm \equiv m_A/(2\gamma (1\mp\beta))$. The width of this box-shaped spectrum is $\Delta E = E_+ - E_- = \sqrt{E_{cm}^2/4-m_A^2} \approx \sqrt{m_\chi^2-m_A^2}$, where we have used $E_{cm}\approx 2m_\chi$. Note that in the special case in which $m_\chi \sim m_A$, the decaying particle is produced at rest, and the box-shaped spectrum reduces to a gamma-ray line at energy $E=m_A/2$.

The spectra from FSR depend on the s-channel mediator. For vector and scalar mediators, they are given respectively by \cite{Coogan:2019qpu}
\begin{align}\label{eq:FSR_Zp}
    \frac{dN^{l^-l^+}}{dE}(E)\Big|_{Z'} = \frac{\alpha_{em}}{E \pi \sqrt{1-4\mu^2}(1+2\mu^2)}\Big[ & (1+(1-x)^2-4\mu^2(x+2\mu^2))\log\left(\frac{1+\sqrt{1-\frac{4\mu^2}{1-x}}}{1-\sqrt{1-\frac{4\mu^2}{1-x}}}\right) \nonumber \\
    & -(1+(1-x)^2 +4\mu^2(1-x) )\sqrt{1-\frac{4\mu^2}{1-x}}
    \Big] 
\end{align}
and 
\begin{align}\label{eq:FSR_hp}
    \frac{dN^{l^-l^+}}{dE}(E)\Big|_{h'} = \frac{\alpha_{em}}{E \pi (1-4\mu^2)^{3/2}}\Big[ & (2(1-x-6\mu^2)+(x+4\mu^2)^2)\log\left(\frac{1+\sqrt{1-\frac{4\mu^2}{1-x}}}{1-\sqrt{1-\frac{4\mu^2}{1-x}}}\right) \nonumber \\
    & -2(1-4\mu^2)(1-x)\sqrt{1-\frac{4\mu^2}{1-x}}
    \Big] 
\end{align}
where we use the dimensionless parameters $x=2E/E_{cm}$ and $\mu = m_l/E_{cm}$.

Finally, the spectrum from cascade FSR is found after boosting the direct FSR spectrum via the corresponding mediator $i$ (Eq.\,\ref{eq:FSR_Zp} or Eq.\,\ref{eq:FSR_hp}) from the rest frame of a decaying particle $A$ to the DM frame \cite{Elor:2015tva}:
\begin{equation}
    \frac{dN^{AB}}{dE}(E)\Big|^{l^-l^+\gamma} = BR(A\to l^-l^+) \frac{4}{E_{cm}}\int_{x_{min}}^{x_{max}} \frac{dx_0}{x_0\sqrt{1-\epsilon_A^2}} \frac{dN^{l^-l^+}}{dx_0}(x_0)\Big|_i\,,
\end{equation}
where the variable $x_0 = 2E_0/E_{cm}$, with $E_0$ the photon energy in the center of mass of the direct annihilation process, is integrated from $x_{min} = \frac{4E}{E_{cm}\epsilon_A^2}(1-\sqrt{1-\epsilon_A^2})$ to $x_{max} = \min\left[ 1, \frac{4E}{E_{cm}\epsilon_A^2}(1+\sqrt{1-\epsilon_A^2}) \right]$ and the variable $\epsilon_A \equiv \frac{2m_A}{E_{cm}}$ parametrizes the boost.

The class of models considered in this work, in which the mediators $h'$ and $Z'$ can have masses close to or lower than the dark matter mass scale, enables a rich phenomenology with gamma rays. As depicted in Fig.\,\ref{fig:diagrams}, we have gamma-ray lines from $\chi \bar \chi \to \gamma \gamma, h' \gamma, \pi^0 \gamma, Z' \gamma$ (panels \textit{a}-\textit{c}); box-shaped spectra from $\chi \bar \chi \to h' \gamma, \pi^0 \gamma, h'h', h'Z'$ (panels \textit{b} and \textit{d}-\textit{f}); cascade FSR from $\chi \bar \chi \to h' h', Z' Z', h' Z'$ (panels \textit{d}-\textit{f}); and FSR from $\chi \bar \chi \to e^- e^+, \mu^- \mu^+$ (panel \textit{g}).

We have implemented our generic vector-scalar model in the numerical package {\tt Hazma} \cite{Coogan:2019qpu,Coogan:2022cdd} by coding the general expressions for the cross-sections of the processes of Fig.\,\ref{fig:diagrams} and the above expressions for the spectra (see Appendix \ref{sec:Hazma}).

In Fig.\,\ref{fig:spectra}, we show the gamma-ray spectra for representative sets of the free parameters. We have assumed the $U(1)_{A}$ case, but the qualitative features are common to vector-scalar portals in general. The gray dashed curves correspond to the spectra seen by a gamma-ray telescope with an energy resolution of $1\%$ FWHM. The blue and red curves are the spectra that could be measured respectively by COSI (covering the energy range $0.2 - 5$ MeV with energy resolution $0.1\%-1.4\%$ FWHM \cite{Beechert:2022phz}) and AMEGO-X (covering the energy range $0.1 - 1000$ MeV with required energy resolution of $5\%$ FWHM at 1 MeV and $17\%$ FWHM at 100 MeV \cite{Caputo:2022xpx}). In the \textbf{upper left panel}, we have $m_{h'}<m_\chi<m_e, m_{Z'}$, so the available final states are $\gamma\gamma, h'\gamma$, and $h'h'$. We can observe the box-shaped contribution from the $h'h'$ channel for $0.07 < E_\gamma/MeV < 0.33$ and a strong gamma-ray line at $E_\gamma = m_\chi$ from $\gamma\gamma$. We also note a small contribution from the $h'\gamma$ channel (a box and a line), which has a smaller annihilation branching fraction. In the \textbf{upper right panel}, we have the special case in which $m_\chi=m_{h'}$, so the $h'h'$ box-shaped spectrum reduces to a gamma-ray line at $E_\gamma=m_\chi/2$. In the \textbf{lower left panel}, we have $m_{h'}, m_e<m_\chi< m_{Z'}$, so DM can also annihilate into $e^-e^+$. In this case, the FSR contribution dominates the spectrum at lower energies. We then note the small contribution from the $h'h'$ decaying to $\gamma\gamma$ and a weaker gamma-ray line, since the annihilation branching fraction is now reduced. In the \textbf{lower right panel}, DM is heavier than the electron and the dark sector particles, in which case the $h' Z'$ channel dominates over the $e^-e^+$ contribution and the gamma-ray line from $\gamma\gamma$ is negligible. Note that an excellent energy resolution is crucial to distinguish the different spectral features allowed by this minimal yet rich class of models. 

\begin{figure}[t!]
    \centering
\includegraphics[width=0.496
\textwidth]{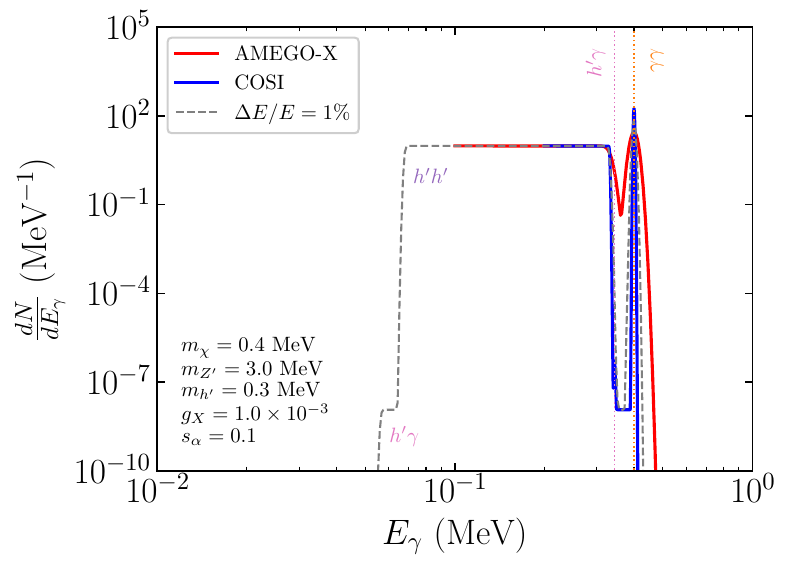} 
\includegraphics[width=0.496
\textwidth]{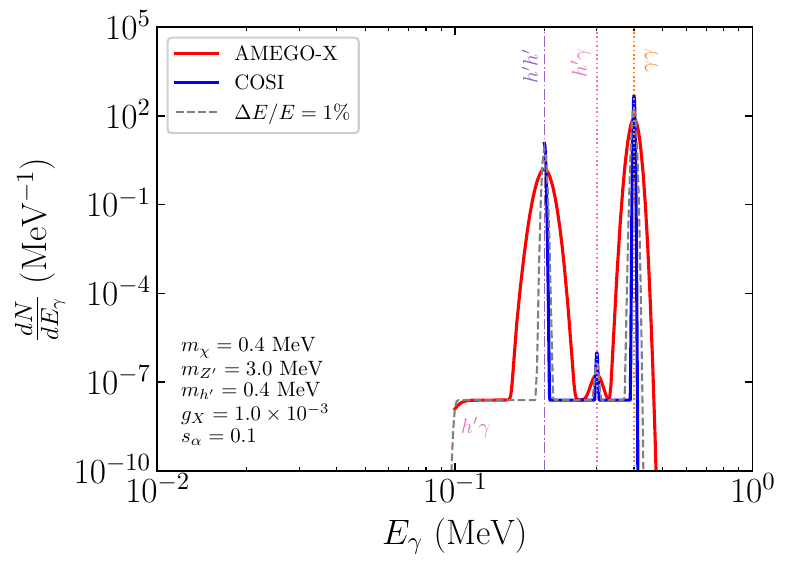}
\includegraphics[width=0.496
\textwidth]{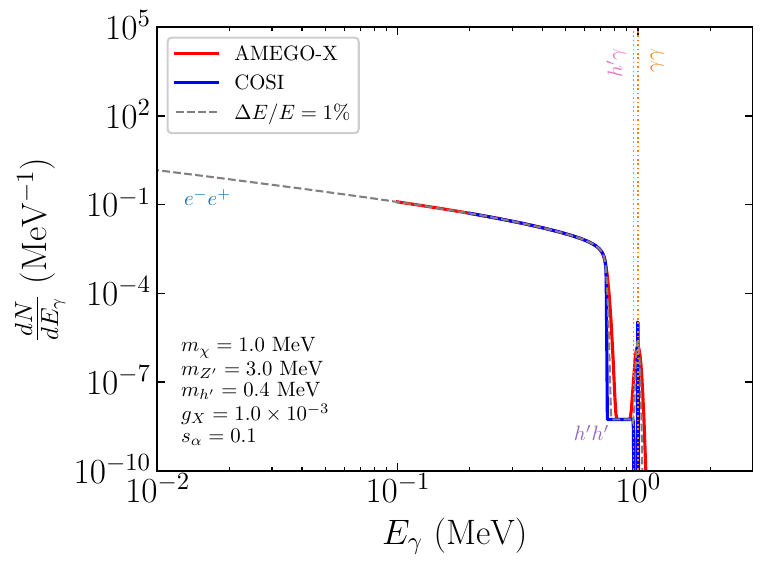} 
\includegraphics[width=0.496
\textwidth]{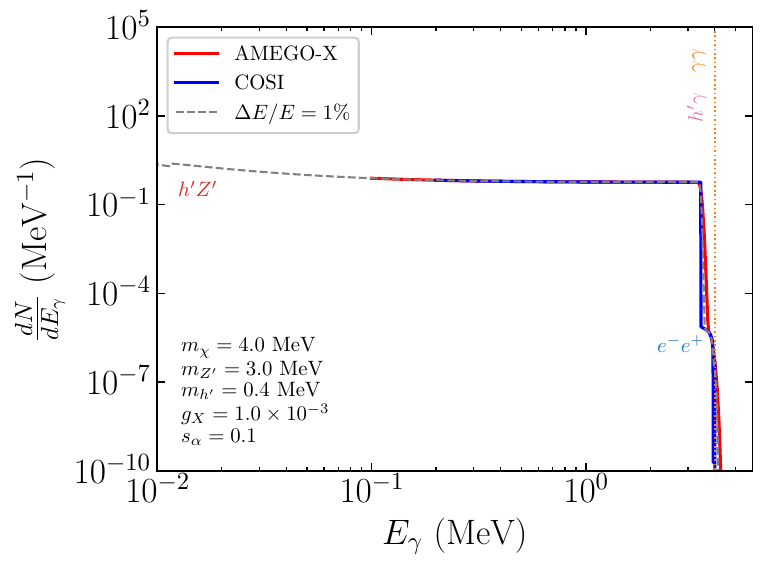}
\caption{Gamma-ray spectra emitted by the annihilation of our light dark matter for different mass hierarchies: $m_{h'}<m_\chi<m_e, m_{Z'}$ (upper left), $m_{h'}=m_\chi<m_e, m_{Z'}$ (upper right), $m_{h'}, m_e<m_\chi< m_{Z'}$ (lower left), and $m_{h'}, m_e, m_{Z'} <m_\chi$ (lower right). This figure shows the possible spectral features in vector-scalar portals as would be seen by the upcoming COSI telescope (blue) and the proposed telescope AMEGO-X (red), as well as by a telescope with a fixed energy resolution of $1\%$ FWHM (dashed gray). }
\label{fig:spectra}
\end{figure}

We computed the cross-sections for the processes depicted in Fig.\,\ref{fig:diagrams} using the {\tt FeynCalc} \cite{Shtabovenko:2023idz} software. We provide below approximate expressions for the most relevant velocity-averaged annihilation cross-sections to help in the interpretation of our results. 

The annihilation into $\gamma\gamma$ and $l^- l^+$ are s-wave through the exchange of $Z'$ and p-wave through the exchange of scalars:
\begin{align}
    \langle \sigma v \rangle_{\gamma \gamma} \approx & \frac{g_X^4 c_\xi^2 A_\chi^2}{64\pi^5}\frac{m_\chi^2}{m_{Z'}^4}\frac{(4-r_{Z'}^2)^2}{(4-r_{Z'}^2)^2+r_{Z'}^2\gamma_{Z'}^2} \Big(\sum_f A_f^X N^c_f e^2 Q_f^2 |F_1^f-F_2^f| + 8\pi^2 \frac{m_\chi^2}{\Lambda^2} \Big)^2 \nonumber \\
 & + v_{rel}^2 \, \frac{s_\alpha^2 c_\alpha^2 \alpha_{em}^2 G_f}{32\sqrt{2}\pi^3}\frac{m_\chi^2}{v_s^2}\frac{|F(m_{h'})|^2}{(4-r_{h'}^2)^2+r_{h'}^2\gamma_{h'}^2}    
\end{align}
and
\begin{align}\label{eq:sv_ee}
    \langle \sigma v \rangle_{l^- l^+} \approx & \frac{c_\xi^2 g_X^2 m_\chi^2}{2\pi m_{Z'}^4} \frac{ \sqrt{1-r_f^2} \Big[ 
    V_{fZ'}^2  V_\chi^2 r_{Z'}^4(2+r_f^2) + 
    A_{fZ'}^2  (2r_{Z'}^4 V_\chi^2(1-r_f^2) + A_\chi^2r_f^2(4-r_{Z'}^2)^2) \Big]}{(4-r_{Z'}^2)^2+r_{Z'}^2\gamma_{Z'}^2}
    \nonumber \\
 & + v_{rel}^2 \, \frac{s_\alpha^2 c_\alpha^2 m_f^2}{8\pi^2 v^2 v_s^2} \frac{(1-r_f^2)^{3/2}}{(4-r_{h'}^2)^2+r_{h'}^2\gamma_{h'}^2}\,,
\end{align}

We have defined the dimensionless variables $r_i \equiv m_i/m_\chi$, $\gamma_i \equiv \Gamma_i/m_\chi$, with $m_i$ and $\Gamma_i$ the mass and decay width of a particle $i$. $G_f$ is the Fermi constant and the interaction strengths between the SM fermions and $Z'$, $V_{fZ'}$ and $A_{fZ'}$, are defined from Eq.\,\ref{eq:Lfermions}.

Note that in the special cases in which $m_{Z'}\approx 2m_\chi$ and $m_{h'}\approx 2m_\chi$, the s-channel mediators are produced on-shell and the cross-sections are resonantly enhanced, except for the $\gamma\gamma$ final state mediated by $Z'$, since a massive spin-1 field cannot decay into two photons.

The annihilation into $h'Z'$ receives comparable contributions from the s-channel, the t-channel, and their interference. This is an s-wave process:
\begin{align}
    \langle \sigma v \rangle_{h' Z'} \approx & \frac{c_\xi^2 g_X^2}{32\pi} \sqrt{1-(r_{h'}+r_{Z'})^2} \sqrt{1-(r_{h'}-r_{Z'})^2} \nonumber \\
    & \times \Big\{ \frac{c_\alpha^2 m_\chi^2/(v_s^2 m_{Z'}^2)}{(4-r_{h'}^2-r_{Z'}^2)^2} \Big[ V_\chi^2 (2(4-r_{h'}^2)^2+r_{h'}^2 r_{Z'}^2(12+r_{h'}^2)-2 r_{Z'}^2(3+r_{h'}^2) +r_{Z'}^6)\nonumber\\
    & \hspace{4.5cm} + A_\chi^2 (2+r_{Z'}^2)((4-r_{Z'}^2)^2-2r_{h'}^2(4+r_{Z'}^2)+r_{h'}^4) \Big] \nonumber \\
    & \hspace{.8cm} + \frac{\lambda_{h'Z'Z'}^2 m_\chi^2 v_s^2/m_{Z'}^6}{32 s_W^4 c_W^4 ((4-r_{Z'}^2)^2+r_{Z'}^2 \gamma_{Z'}^2)}\Big[ V_\chi^2 r_{Z'}^4 ( (4-r_{h'}^2)^2 + 2r_{Z'}^2(20-r_{h'}^2) + r_{Z'}^4 ) \nonumber \\
    & \hspace{4.5cm} + A_\chi^2 (4-r_{Z'}^2)^2 ((4-r_{Z'}^2)^2 - 2r_{h'}^2(4+r_{Z'}^2)+r_{h'}^4) \Big] \nonumber \\
    & \hspace{.8cm} - \frac{c_\alpha \lambda_{h'Z'Z'} m_\chi^2/m_{Z'}^4 (4-r_{Z'}^2)}{2s_W^2c_W^2( (4-r_{Z'}^2)^2 + r_{Z'}^2 \gamma_{Z'}^2 )(4-r_{h'}^2-r_{Z'}^2)} \nonumber \\
    & \hspace{1.5cm} \times \Big[ V_\chi^2 r_{Z'}^2 ( (4-r_{h'}^2)^2 + 4r_{Z'}^2(4+r_{h'}^2)-5r_{Z'}^4 ) \nonumber \\
    & \hspace{2cm} + A_\chi^2(4-r_{Z'}^2)( (4-r_{Z'}^2)^2 - 2r_{h'}^2(4+r_{Z'}^2)+r_{h'}^4 ) \Big] \Big\}
\end{align}

Finally, the annihilation into $h'h'$ receives comparable contributions from the $s-$, $t-$, and $u-$channels and from their interferences. This is a p-wave process:
\begin{align}
    \langle \sigma v \rangle_{h' h'} \approx & v_{rel}^2 \frac{c_\alpha^2 \sqrt{1-r_{h'}^2}}{384\pi^2} \left[ \frac{27 \pi \lambda_{h'h'h'}^2}{(4-r_{h'}^2)^2+r_{h'}^2 \gamma_{h'}^2} + \frac{8c_\alpha^2 m_\chi^2/v_s^4}{(2-r_{h'}^2)^2} + \frac{12c_\alpha \lambda_{h'h'h'}(4-5r_{h'}^2+r_{h'}^4)}{(2-r_{h'}^2)^2 ((4-r_{h'}^2)^2+r_{h'}^2 \gamma_{h'}^2)} \right]
\end{align}

In the expressions above, we defined the coupling strengths between mediators $\lambda_{h'Z'Z'} = s_\alpha (e s_\xi-2c_\xi s_W c_W g_X X_\Phi)^2 - c_\alpha (2 c_\xi s_W c_W g_X X_{\Phi_S})^2$ and $\lambda_{h'h'h'} = -2 c_\alpha^3 \lambda_S - c_\alpha s_\alpha^2 \lambda_{HS} + (2 s_\alpha^3 \lambda + s_\alpha c_\alpha^2 \lambda_{HS}) \frac{v}{v_s}$.

\begin{figure}[t!]
    \centering
\includegraphics[width=0.98
\textwidth]{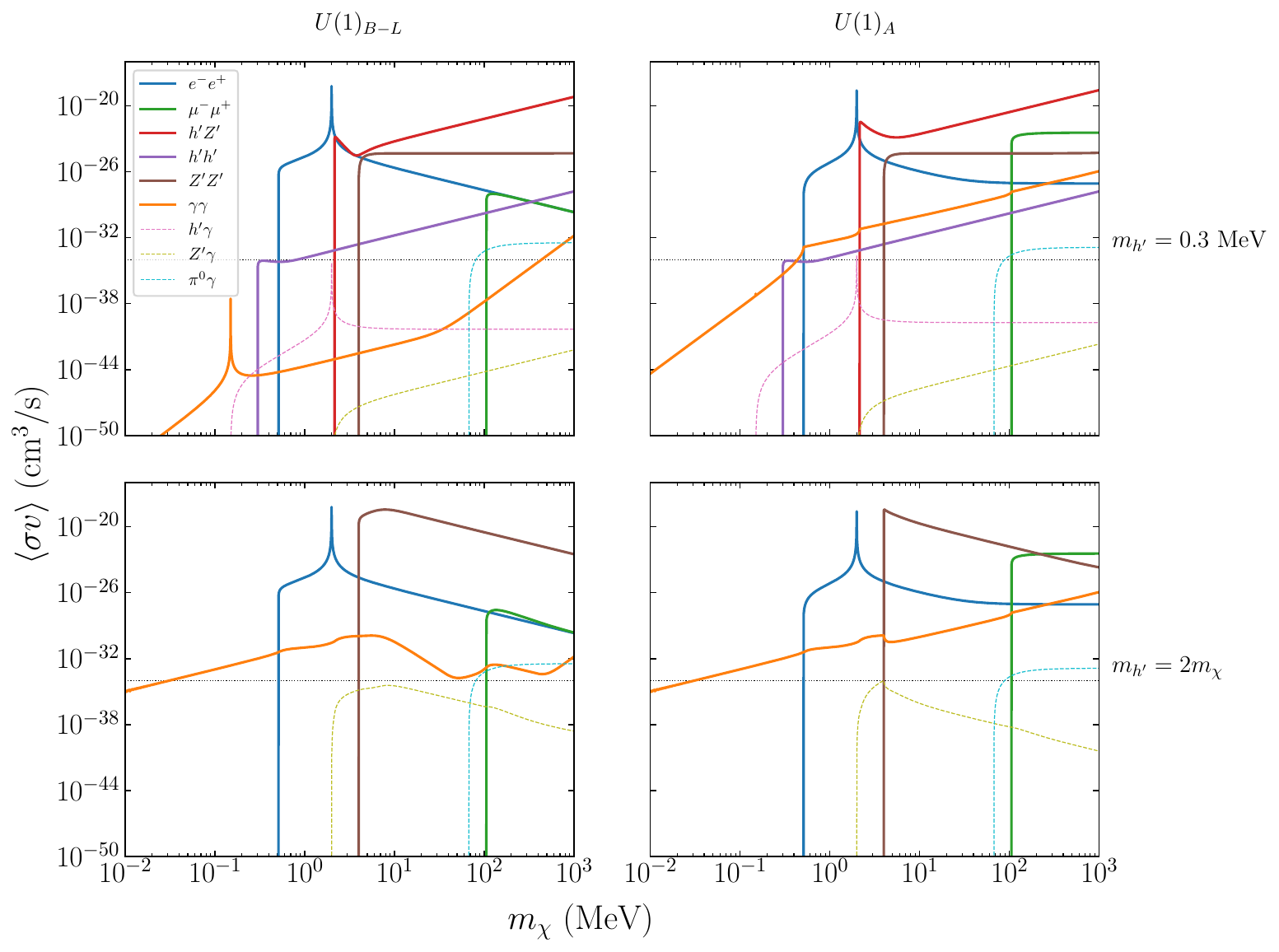} 
\caption{Velocity-averaged annihilation cross-section of our light dark matter into channels leading to $\gamma$ rays for different charge assignments ($U(1)_{B-L}$ in left panels and $U(1)_A$ in right panels) and different values of singlet scalar mass ($m_{h'} = 0.3$ MeV in top panels and $m_{h'} = 2 m_\chi$ in bottom panels). The remaining parameters are set to $g_X = 10^{-3}$, $s_\alpha = 0.1$, and $m_{Z'}=4$ MeV. The annihilation into leptons is dominated by the exchange of $Z'$ bosons, which are produced on-shell when $m_\chi = m_{Z'}/2$. This leads to the resonance enhancement observed in the $e^+ e^-$ channel. In the $U(1)_{B-L}$ case, the $\gamma \gamma$ channel is relevant only in the regime of singlet scalar resonance (bottom left), whereas in the $U(1)_A$ case, it is relevant provided $g_X$ is strong enough and for light enough $Z'$, independently of the scalar mixing.
}
\label{fig:CrossSections}
\end{figure}

In Fig.\,\ref{fig:CrossSections}, we show the contributions of the different channels to the total velocity-averaged annihilation cross-section. In the left panels, we have the $U(1)_{B-L}$ case, and in the right panels, we have the $U(1)_A$ case. In the upper panels, we have the singlet scalar lighter than the electron ($m_{h'}=0.3$ MeV). As we will show later, in this case the $h'h'$ channel becomes more relevant. In the lower panels, we show the special case of scalar resonance ($m_{h'}=2m_\chi$). The remaining parameters are set to $g_X = 10^{-3}$, $s_\alpha = 0.1$, and $m_{Z'}=4$ MeV. The horizontal line indicates the scale of the cross-section that gamma-ray line searches could constrain ($\langle \sigma v \rangle \sim 10^{-34}$ cm$^3/$s), as we will show later.

Because of the Yukawa suppression in the DM-scalar interactions ($\sigma \propto m_\chi^2/v_s^2$), the annihilation into leptons is dominated by the exchange of $Z'$ bosons. When $m_\chi = m_{Z'}/2 \approx 2$ MeV, $Z'$ bosons are produced on-shell before decaying into leptons. This leads to the resonance enhancement observed in the $e^+ e^-$ channel in all the panels. The channel $h'Z'$ happens mainly through the s-channel exchange of $Z'$ near threshold and through t-channel exchange of $\chi$ far from the threshold. It easily dominates the total cross-section whenever available (as in the upper panels). Regarding the $\gamma\gamma$ channel, which leads to the strongest MeV gamma-ray lines, it dominates the total cross-section when $m_\chi < m_e$, as expected. For $U(1)_{B-L}$, gamma-ray lines would be only detectable in the special region of singlet scalar resonance, where $m_{h'} \approx 2m_\chi$. In this case, the $Z'Z'$ channel, leading to cascade FSR, and the $Z'\gamma$, leading to gamma-ray lines and cascade FSR, are also strongly enhanced. It is worth mentioning that for $m_\chi \gtrsim 10 m_{h'}$, there is an enhancement in the $\gamma\gamma$ cross-section due to the contribution of the BSM fermions in the loop. We have set $\Lambda=1$~TeV as a benchmark value. Thus, we would also have detectable gamma-ray lines in the $U(1)_{B-L}$ case for very light singlet scalar and/or lower values of $\Lambda$. The $U(1)_A$ case is much more promising for gamma-ray lines. In this case, the annihilation is dominated by $Z'$ exchange, so it is significantly enhanced (non-resonantly) for lighter $Z'$ but quickly weakens for lower values of $g_X$ ($\sigma \propto g_X^4 m_\chi^2/m_{Z'}^4$).

Finally, we note that in vector-scalar portal models where all the SM particles are neutral under $U(1)_X$, we can still expect strong limits from gamma-ray telescopes. In this case, the situation for the $\gamma\gamma$ channel is similar to the $U(1)_{B-L}$ case, as the scalar couplings are not affected by the gauge charges. The annihilation into leptons happens only via scalar exchanges and becomes significantly weakened, well below the detectability scale. However, the $h'h'$ and the $h'Z'$ channels are not affected, so we could constrain this class of models by searching for continuum gamma-rays, possibly with a box-shaped spectrum.

\section{Indirect detection constraints}
\label{sec:ID}

The current gamma-ray limits on the annihilation of MeV-scale dark matter are set by COMPTEL and INTEGRAL/SPI from measurements of the diffuse gamma-ray spectrum, which include contributions from the Galactic diffuse continuum emission (containing bremsstrahlung and inverse Compton scatterings), gamma-ray point sources in their field-of-view, and extra-galactic diffuse emission. 

We use the diffuse gamma-ray spectrum measured by COMPTEL at energies from 1 to 30 MeV in the search region $|b| < 5\degree, |l| < 30\degree$ ($\Delta \Omega \simeq 0.18$ sr) \cite{Strong:1998ck}. The energy resolution of COMPTEL can be fitted as $\sigma(E) = 0.01 (14.61 E/MeV + 2.53 (E/MeV)^2)^{1/2}$ \cite{Schoenfelder1993}. For the INEGRAL/SPI limits, we use the diffuse gamma-ray emission at energies from 0.02 to 2.4 MeV in the search region of $|b|,|l| < 47.5\degree$ ($\Delta \Omega = 2.44$ sr) \cite{Siegert:2022jii,Berteaud:2022tws}. The energy resolution of INTEGRAL/SPI is about $0.2\%$ FWHM at 1 MeV~\footnote{\url{https://integral.esac.esa.int/AO21/SPI_ObsMan.pdf}.}. The 46-day COSI balloon flight of 2016, done in preparation for the COSI telescope, measured the diffuse emission at energies from 0.2 to 5 MeV in the search region $|b| < 45\degree, |l| < 65\degree$ ($\Delta \Omega \simeq 3.21$ sr) \cite{COSI:2023bsm}. The energy resolution calibrated for the 2016 COSI balloon flight is fitted as $\epsilon = 0.31 (E/MeV)^{-0.95}$ FWHM \cite{Beechert:2022phz}. 

Many MeV gamma-ray telescopes are currently being built, such as the COSI telescope \cite{Tomsick:2023aue}, and planned, such as AMEGO-X \cite{Caputo:2022xpx}. The narrow-line and continuum point-source sensitivities projected for the COSI telescope were reported in Ref.\,\cite{Tomsick:2023aue}. These are based on the mission requirements at the 3-$\sigma$ level for 2 years of survey. The point-source 3-$\sigma$ sensitivity projected for 3 years of the AMEGO-X mission was reported in Ref.\,\cite{Caputo:2022xpx}. Following Refs.\,\cite{Negro:2021urm,Caputo:2022dkz}, we have done a rescaling of these point-source sensitivities to account for an extended target region. Specifically, we multiply the point-source sensitivity in each energy bin by the factor $(\Delta\Omega/\Delta\Omega_{res})^{1/4}$, where $\Delta\Omega$ is the source extension and $\Delta\Omega_{res}=\pi R_{ARM/2}^2$ quantifies the angular resolution of the Compton telescope, with $R_{ARM/2}$ the half width at half maximum of the angular resolution measure distribution. We consider as the target region for our estimated prospects a $10\degree$ disk around the GC ($\Delta\Omega = 0.095$ sr) and perform the scaling considering the required angular resolution for the COSI mission \cite{Tomsick:2023aue} and the simulated angular resolution for AMEGO-X \cite{Caputo:2022xpx}.

\begin{figure}[t!]
    \centering
\includegraphics[width=0.6
\textwidth]{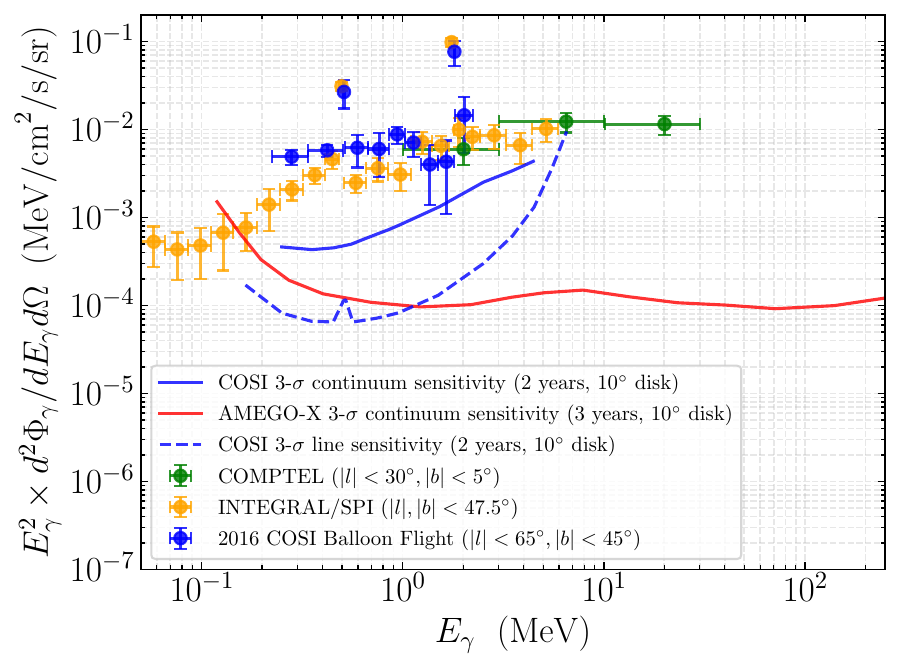} 
\caption{Measurements of diffuse gamma-rays from COMPTEL \cite{Strong:1998ck}, INTEGRAL/SPI \cite{Siegert:2022jii,Berteaud:2022tws}, and the COSI balloon flight of 2016 \cite{COSI:2023bsm}, as well as the scaled point-source continuum (blue solid) and line (blue dashed) sensitivities reported for the COSI mission \cite{Tomsick:2023aue} and the scaled point-source continuum sensitivity for AMEGO-X \cite{Caputo:2022xpx} (red). }
\label{fig:fluxes}
\end{figure}

In Fig.\,\ref{fig:fluxes}, we show the data points for COMPTEL (green), INTEGRAL/SPI (orange), and the 2016 COSI balloon flight (blue), as well as the scaled sensitivities of the COSI telescope in the case of continuum (solid blue curve) and line (dashed blue curve) searches, and of the AMEGO-X continuum searches (red curve), averaged by $\Delta \Omega$.

The annihilation of MeV DM into electromagnetic particles between the periods of recombination and ionization can significantly affect the temperature and polarization power spectra of the CMB and is thus subject to a strong constraint from the measurements of the Planck satellite \cite{Planck:2018vyg}:
\begin{equation}
    f_{eff}\frac{\langle \sigma v \rangle_{CMB}}{m_\chi} < 3.5 \times 10^{-31} \text{cm}^3/\text{s}/\text{MeV}\,,
\end{equation}
where $\langle \sigma v \rangle_{CMB}$ is the velocity-averaged annihilation cross-section at the time of the CMB release and $f_{eff}$ quantifies the efficiency of the energy deposition from DM annihilation in the medium \cite{Slatyer:2012yq,Slatyer:2015jla,Slatyer:2015kla}. In our case, $f_{eff}$ receives contributions of the total photon spectrum $dN/dE_\gamma$ and twice the total positron spectrum $dN/dE_{e^+}$, from the electron-positron pair produced per DM annihilation:
\begin{equation}
   f_{eff} = \frac{1}{E_{cm}}\int_0^{E_{cm}/2} dE \,E \left( f_{eff}^\gamma \frac{dN}{dE_\gamma}(E) + 2 f_{eff}^{e^+} \frac{dN}{dE_{e^+}}(E) \right) \,.
\end{equation}

In the expression above, $E$ is the energy of the photon or the positron, and $f_{eff}^{\gamma,e^+}$ are the coefficients defined in Ref.\cite{Slatyer:2015jla}. The computation of the positron spectrum is done similarly to the gamma-ray case, with positron lines coming from $\chi\bar\chi\to e^-e^+$, at $E=E_{cm}/2$, cascade annihilation from the muon decay from $\chi\bar\chi\to \mu^+\mu^-$ followed by $\mu^+\to e^+ \nu_e \bar\nu_\mu$, and cascade annihilation from $h'$ and $Z'$ decays into $e^+e^-$ ($\chi\bar\chi\to h'\gamma, Z'\gamma, h'h', Z'Z', h'Z'$).

When the DM annihilation depends on velocity (p-wave), we must relate $\langle \sigma v \rangle_{CMB}$ to today's $\langle \sigma v \rangle$: $\langle \sigma v \rangle = \langle \sigma v \rangle_{CMB} (v_0/v_{CMB})^2$. The velocity of a thermal relic at the CMB time is given in terms of the DM temperature $T_{DM}$: $v_{CMB}=\sqrt{3 T_{DM}/m_\chi}$, with $T_{DM}=T_{kd}(z/z_{kd})$ after kinetic decoupling at redshift $z_{kd}$. Since $T_\gamma \propto 1/a = 1+z \approx z$, we have
\begin{equation}
    v_{CMB} \approx 2\times 10^{-4} \frac{T_\gamma}{1 \text{eV}} \frac{1\text{MeV}}{m_\chi}\sqrt{\frac{10^{-4}}{x_{kd}}}\,,
\end{equation}
with $x_{kd}=T_{kd}/m_\chi$. Even though the value of $x_{kd}$ is model-dependent, for a typical WIMP dark matter, the kinetic decoupling happens in the non-relativistic regime after freeze-out and $x_{kd}\ll 1$. 

One of the main challenges of the indirect detection of sub-GeV dark matter is the theoretical uncertainty regarding the calculation of gamma-ray and positron spectra. At center-of-mass energies below a few GeV, the quarks and gluons produced from DM annihilation hadronize and cannot be treated as parton showers as assumed in the {\tt PYTHIA} simulations \cite{Bierlich:2022pfr} used in numerical packages for DM phenomenology such as {\tt micrOMEGAs} \cite{Alguero:2023zol}. The python toolkit {\tt Hazma} \cite{Coogan:2019qpu} was recently introduced to deal with this challenge, providing a framework to derive indirect detection and CMB limits on sub-GeV DM.

In the class of models we are considering, in which DM can annihilate into leptons and sub-GeV mediators, hadronic final states are not expected to dominate the spectrum\,\footnote{See e.g. Ref.\,\cite{Boddy:2015efa} for models where light mesons can dominate the annihilation of light DM.}. Nevertheless, we include in our study the annihilation channel $\pi^0\gamma$ as it is a source of gamma-ray lines and boxes potencially detectable by AMEGO-X. For the purposes of our work, we neglect DM annihilation into two pions and heavier mesons.

In the current version of the {\tt Hazma} code, the effective interactions between DM and light hadrons in the built-in DM models are derived using chiral perturbation theory at leading order \cite{Coogan:2021sjs}. However, user-defined models assuming alternative hadronic treatments are possible. We have adopted the Vector Meson Dominance approach for the $Z'\pi^0\gamma$ vertex \cite{Tulin:2012uq}. As it turns out, the process $\bar\chi\chi\to \pi^0\gamma$ is always subleading in the class of models considered in this work (see Fig.\,\ref{fig:CrossSections}) and uncertainties in the hadronic modeling are not expected to impact the indirect detection limits in our case. 

We used the numerical package {\tt Hazma} to derive the indirect detection limits from MeV gamma-ray and CMB observations. We updated our version of the code with the specific target information, as well as the energy resolution and data for the telescopes considered in this work, as specified above. For the gamma-ray observations, we have used the binned analysis of {\tt Hazma}. Specifically, we compare the total gamma-ray flux from DM annihilation $\frac{d\Phi}{dE_\gamma} (E_\gamma, \Delta \Omega) \Big|_{\bar \chi \chi}$ (Eq.\,\ref{eq:fluxDM}), convolved with the telescope energy resolution $\epsilon$ and integrated over the energy range of the bin, to the largest total flux measured/measurable in each bin, ${\Phi_\gamma^{obs}\Big|}_{max}^{(i)} = {\Phi_\gamma^{obs}\Big|}^{(i)} + n_\sigma \sigma^{(i)}$, where $n_\sigma$ is the significance of the upper error on the flux $\sigma^{(i)}$. By finding the smallest cross-section for which ${\Phi_\gamma^{\chi\chi}\Big|}_\epsilon^{(i)} > {\Phi_\gamma^{obs}\Big|}_{max}^{(i)}$, we set an upper limit on $\langle \sigma v \rangle$ for a given DM mass. For the COSI and AMEGO-X prospects, we have binned and rescaled the reported point-source sensitivities assuming no errors in the flux values.

\begin{figure}[t!]
    \centering
\includegraphics[width=0.496
\textwidth]{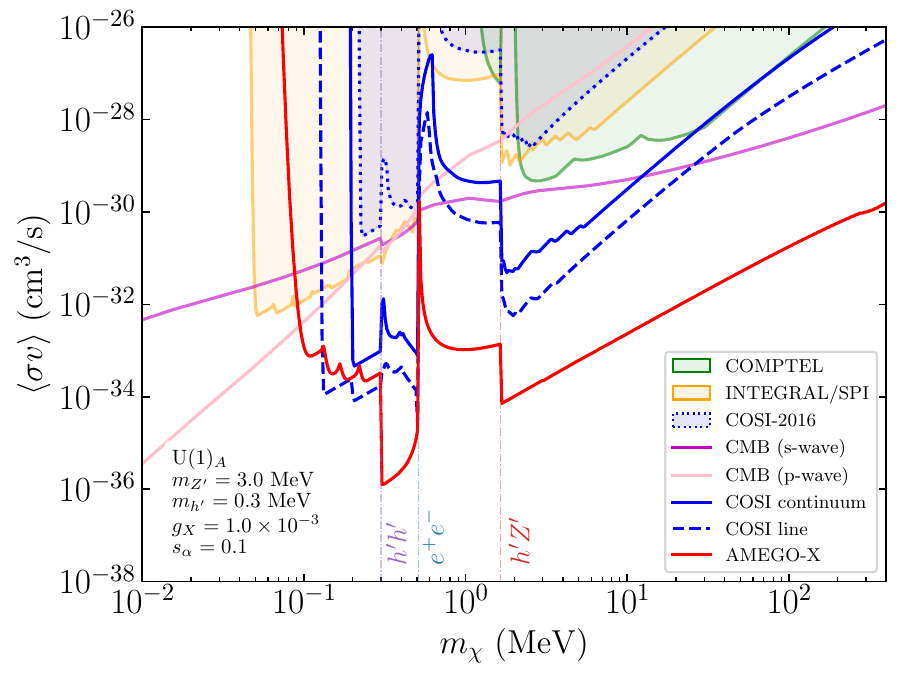} 
\includegraphics[width=0.496
\textwidth]{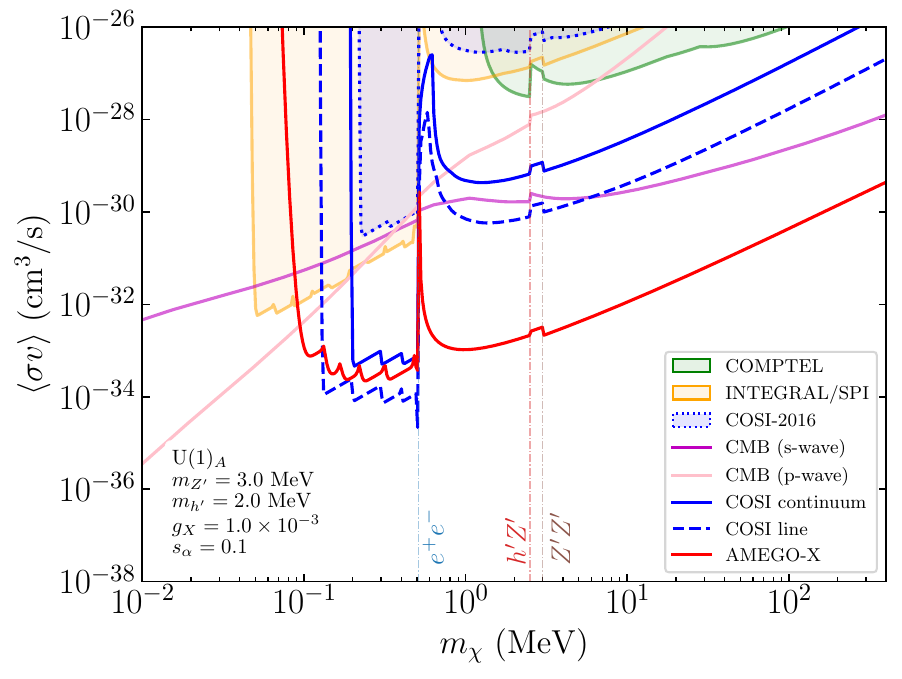}
\caption{Indirect detection constraints on the total dark matter annihilation cross-section as a function of the dark matter mass, for $m_{h'}<m_e$ (left panel) and $m_{h'}>m_e$ (right panel). The colored regions are excluded by COMPTEL, INTEGRAL/SPI, and the 2016 COSI balloon flight (green, orange, and blue regions). The regions above the magenta and pink curves are excluded by s-wave and p-wave CMB limits, respectively. The regions above the dashed and solid blue curves will be probed by COSI (line and continuum gamma-ray searches, respectively) and the regions above the red curve will be probed by AMEGO-X. The vertical lines are the thresholds of the different channels, color-coded as in Fig.\,\ref{fig:CrossSections}.}
\label{fig:sv-mdm}
\end{figure}

In Fig.\,\ref{fig:sv-mdm}, we show the indirect detection constraints on the total velocity-averaged annihilation cross-section as a function of DM mass. The colored regions are already excluded by the gamma-ray telescopes COMPTEL (green), INTEGRAL/SPI (yellow), and the 2016 COSI balloon flight (blue). The regions above the dashed and solid blue curves will be probed by line and continuum searches with the COSI telescope, respectively, and the region above the red curve will be probed by AMEGO-X. As we have discussed, the CMB limits are model-dependent. We show the upper bound on the total cross-section in the limiting cases where s-wave processes dominate (magenta) and where p-wave processes dominate (pink), with $x_{kd}=10^{-4}$. The vertical dot-dashed lines indicate the thresholds of the annihilation channels, following the color code of Fig.\,\ref{fig:CrossSections}. To illustrate how the annihilation channels impact our results, we show different mass hierarchies for the singlet scalar. In the left panel, we have $m_{h'}<m_e$ ($m_{h'}=0.3$ MeV) and in the right panel, we have $m_{h'}>m_e$ ($m_{h'}=2$ MeV). The mass of the $Z'$ is also at the MeV scale, $m_{Z'}=3$ MeV, and the other coupling strength parameters are set to $g_X=10^{-3}$ and $s_\alpha = 0.1$. The charge assignment is set for the $U(1)_A$ case, but since these limits depend mainly on the spectral features (see Eq.\,\ref{eq:fluxDM}), we have similar results for the $U(1)_{B-L}$ case. 

For $m_\chi < m_{h'}, m_{Z'}, m_e$, only the $\gamma\gamma$ channel is open and cross-sections as small as $10^{-34} \text{cm}^3$/s will be probed by COSI line searches. When $m_{h'} < m_\chi < m_{Z'}, m_e$, as in the left panel, the continuum box-shaped spectrum from the $h'h'$ channel dominates the flux. The highest sensitivity of AMEGO-X to continuum gamma rays increases the discovery reach by almost two orders of magnitude. For $m_\chi > m_e$, the continuum gamma rays from FSR typically dominate the flux. However, when DM annihilates into singlet scalars lighter than the electron, the box-shaped contribution from $h'h'$ and especially $h'Z'$ dominate the flux, and the sensitivities of the gamma-ray telescopes increase significantly. When $m_{h'}>m_e$ (as in the right panel), the cascade FSR dominates over the boxes as $BR(h'\to e^- e^+) \gg BR(h'\to \gamma\gamma)$.  

\section{Other constraints on the parameter space}
\label{sec:OtherConstraints}

Let us now consider other constraints on our parameter space. In what follows, we will discuss the achievement of the relic abundance, some important theoretical requirements, and the direct detection searches for our sub-GeV dark matter candidate. Limits coming from collider searches for DM can also be relevant in our case. Moreover, the light mediators $Z'$ and $h'$ can modify SM interactions studied at colliders and accelerators in detectable ways. Since no events compatible with such signals have been found, model-dependent limits can be put on the couplings of DM and mediators to SM particles. Recasting the existing constraints on minimal vector portal models for the specific cases of our vector-scalar portal scenario must be done carefully and is beyond the scope of the present study. These limits can be competitive to the COSI limits for $Z'$ masses larger than 1~MeV \cite{Watanabe:2025pvc}. For this reason, we will explore the possible complementarity between the future MeV gamma-ray telescopes and colliders in the context of vector-scalar portal models in a future work.

\subsection{Relic abundance}

The achievement of the correct relic abundance strongly constrains our parameter space. In vector-scalar portal models, DM can be produced and destroyed in the early universe due to its interactions with particles in the thermal bath mediated by $Z'$ and $h'$. For the relatively large couplings considered in this work, both the mediators and the DM were once part of the thermal bath, and the relic abundance is achieved via the freeze-out mechanism. We computed the relic abundance of our dark matter candidate $\chi$ via freeze-out using {\tt micrOMEGAs} \cite{Alguero:2023zol}. As we discuss in detail in Section \ref{sec:results}, the freeze-out is set by annihilation into the mediators in a large region of the parameter space. As a consequence, the thermal-averaged cross-section corresponding to the right amount of relic abundance in the class of models considered in this work is much below the canonical value $\langle \sigma v \rangle \sim 10^{-26}$ cm$^3/$s. 

\subsection{Theoretical constraints}

Our parameter space is subject to two theoretical constraints: vacuum stability of the potential, which ensures that the theory has a stable ground state, and perturbative unitarity, which ensures the conservation of probability at a given order in perturbation theory \cite{Langacker:2017uah,Logan:2022uus,Hiller:2024zjp}. 

In this work, we require that the tree-level scalar potential is bounded from below. In this case, it suffices to ensure $\lambda, \lambda_s >0$ and $|\lambda_{hs}| < 2 \sqrt{\lambda \lambda_s}$. From Eq.\,\ref{Eq:lambdas}, the first two conditions are always valid, whereas the third condition restricts the values of $m_{h'}$ and $s_\alpha$. On the other hand, the perturbative unitarity constraints on the quartic couplings, $\lambda < 4\pi$, $\lambda_s < 4\pi$, and $\lambda_{hs}<4\pi$, translate respectively into the following upper bounds on the singlet scalar mass and gauge coupling after using Eqs.\,\ref{Eq:lambdas} and \ref{eq:vs}:
\begin{align}
  & m_{h'} < \sqrt{8\pi v^2-c_\alpha^2 m_h^2}/s_\alpha \\
  & g_X < \left(X_{\Phi_s}^2 \frac{s_\alpha^2 m_h^2+c_\alpha^2 m_s^2}{8\pi m_{Z'}^2}- X_\Phi^2 \frac{4 m_{Z^0}^2}{g_Z^2 (m_{Z^0}^2-m_{Z'}^2)}\right)^{-1/2} \label{eq:ls} \\
  & g_X < 4\pi\left(X_{\Phi_s}^2\frac{s_\alpha^2 c_\alpha^2 (m_h^2-m_s^2)^2}{m_{Z'}^2 v^2}-X_\Phi^2\frac{64 \pi^2 m_{Z^0}^2}{g_Z^2 (m_{Z^0}^2-m_{Z'}^2)}\right)^{-1/2}
\end{align}

It is also important to emphasize that, since our dark matter candidate must have a non-zero axial coupling to the $Z'$, DM cannot be arbitrarily heavier than the $Z'$ boson. As discussed in Ref.\,\cite{Kahlhoefer:2015bea}, perturbative unitarity can be violated at high energies by the DM self-scattering mediated by a $Z'$, rendering the constraint $g_X c_\xi A_\chi \lesssim \sqrt{\pi/2} m_{Z'}/m_\chi$. Moreover, to restore the unitarity violated by the t-channel annihilation $\bar \chi \chi \to Z' Z'$ at high energies, we must assume the presence of new physical states that cannot be much heavier than DM and the $Z'$ boson. It is well-known that scalar bosons, which give mass to vector bosons, restore unitarity (just like in the SM). Thus, in vector-scalar portal models with non-zero axial couplings, we have an additional unitarity constraint:
\begin{equation}
g_X^2 < \frac{\pi}{c_\xi^2 A_\chi^2} \frac{m_{Z'}^2}{m_\chi m_{h'}} \,.
\end{equation}

\subsection{Direct detection}
\label{Sec:DDcolliders}

\begin{figure}[t!]
    \centering
\includegraphics[width=0.55
\textwidth]{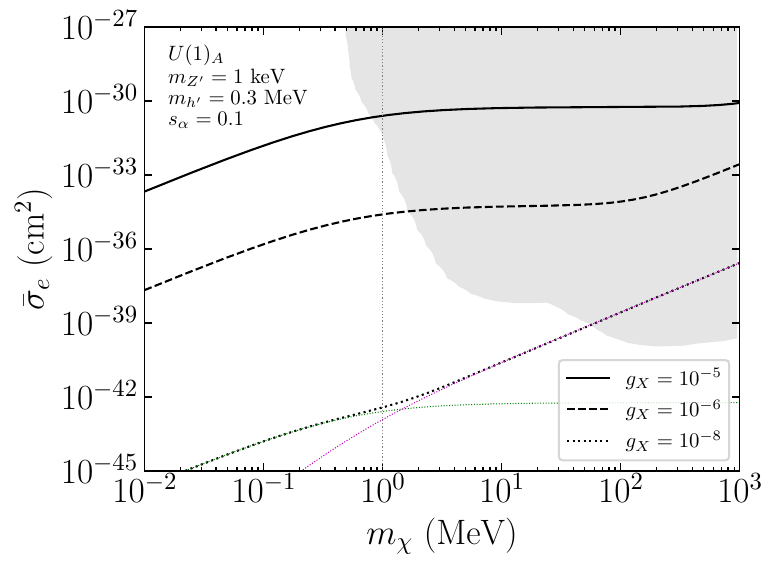} 
\caption{DM-electron elastic scattering cross-section for different values of gauge coupling in the case of $\mathbf{U(1)_{A}}$. For each value of $g_X$, the contribution of the $Z'$ ($h'$) exchange dominates at lower (higher) DM masses and are shown in the case of $g_X=10^{-8}$ in green (magenta). The gray region is already excluded by XENON1T \cite{XENON:2019gfn}, DarkSide-50 \cite{DarkSide:2022knj}, SENSEI \cite{SENSEI:2023zdf}, and DAMIC-M \cite{DAMIC-M:2025luv}. The vertical line at $m_\chi = 1$ MeV is for reference. The complementarity between direct and indirect detection will be possible in the multi-MeV DM regime.}
\label{fig:DD}
\end{figure}

One of the main ways of searching for dark matter is via direct detection experiments. These are underground detectors that could record energy deposition due to DM scattering off nuclei and electrons. Over the last decades, direct detection experiments have focused on the search of nuclear recoils from DM-nucleon elastic scatterings, placing leading constraints on GeV-scale WIMP models \cite{Arcadi:2024ukq}. The direct detection of sub-GeV dark matter is very challenging because the nuclear recoil energy is well below the energy threshold of the detectors. However, many detectable signals of sub-GeV dark matter in underground detectors are being explored \cite{Essig:2011nj, Essig:2015cda,Ibe:2017yqa, Dolan:2017xbu,Essig:2024wtj}. Current experiments already place constraints on the DM-electron cross-section \cite{XENON:2019gfn,DarkSide:2022knj,SENSEI:2023zdf,DAMIC-M:2025luv}, and many experiments are being developed and planned to improve the current sensitivity to the parameter space of sub-GeV dark matter \cite{Battaglieri:2017aum,Knapen:2021run,Essig:2024wtj}.

In vector-scalar models, the cross-section for the elastic scattering off electrons is approximately \cite{Essig:2011nj}
\begin{align}
\bar \sigma_e \approx & \frac{4 g_X^2 c_\xi^2 \mu_{\chi e}^2}{\pi m_{Z'}^4}\left(V_\chi^2 V_{eZ'}^2
+ A_\chi^2 A_{eZ'}^2 \right) + \frac{4 s_\alpha^2 c_\alpha^2 \mu_{\chi e}^2}{\pi m_{h'}^4}\frac{m_e^2 m_\chi^2}{v^2 v_s^2} \nonumber \\
& + 8 s_\alpha c_\alpha g_X c_\xi V_\chi V_{eZ'} \mu_{\chi e}^2
\frac{m_e m_\chi}{\pi v v_s} 
\frac{1+\alpha_{em}^2 m_e/(4m_\chi)}{(m_{Z'}^2-\alpha_{em}^2 m_e^2)(m_{h'}^2-\alpha_{em}^2 m_e^2)}
\,.
\label{eq:DD}
\end{align}

In the expression above, we assume that all masses are much larger than the typical momentum transferred ($m_{Z'},m_{h'},m_\chi \gg \alpha_{em} m_e$). The interaction strengths between the electrons and $Z'$,
$V_{eZ'}$ and $A_{eZ'}$, are defined from Eq.\,\ref{eq:Lfermions}. In the first (second) term, the scattering is dominated by t-channel $Z'$ ($h'$) exchanges, whereas the third term corresponds to the interference among the mediators. The interference is constructive or destructive depending on the mediator masses. The contributions of SM bosons can be safely neglected in the case of light $Z'$/$h'$. 

In Fig.\,\ref{fig:DD}, we show the current constraints on the DM-electron elastic scattering as a function of the DM mass from the direct detection experiments XENON1T \cite{XENON:2019gfn}, DarkSide-50 \cite{DarkSide:2022knj}, SENSEI \cite{SENSEI:2023zdf}, and DAMIC-M \cite{DAMIC-M:2025luv}. The predicted cross-sections are shown for three values of gauge coupling: $g_X = 10^{-5}, 10^{-6}, 10^{-8}$ (solid, dashed, and dotted lines, respectively). They are computed with the charge assignment of $U(1)_A$ and we keep all the $\mathcal{O}(\alpha_{em}^2)$ terms not shown in Eq.\,\ref{eq:DD}. The DM-electron cross-section is dominated by the exchange of $Z'$ bosons for light DM, since this contribution is not suppressed by the Yukawa couplings. The exchange of $h'$ bosons, strongly dependent on the DM mass, dominates for higher DM masses.

\section{Results}
\label{sec:results}

Now that we understand how the different processes contribute to the total annihilation cross-section (Fig.\,\ref{fig:CrossSections}) and how they impact the indirect detection limits (Fig.\,\ref{fig:sv-mdm}), we are in a position to show our main results. Namely, under which circumstances the future MeV gamma-ray telescopes will probe viable regions of the parameter space of vector-scalar portal models. 

\begin{figure}[t!]
    \centering
\includegraphics[width=0.496
\textwidth]{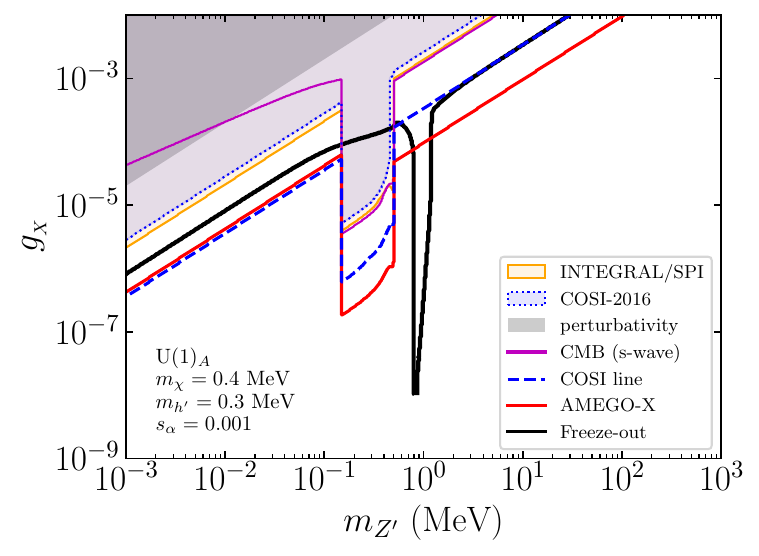} 
\includegraphics[width=0.496
\textwidth]{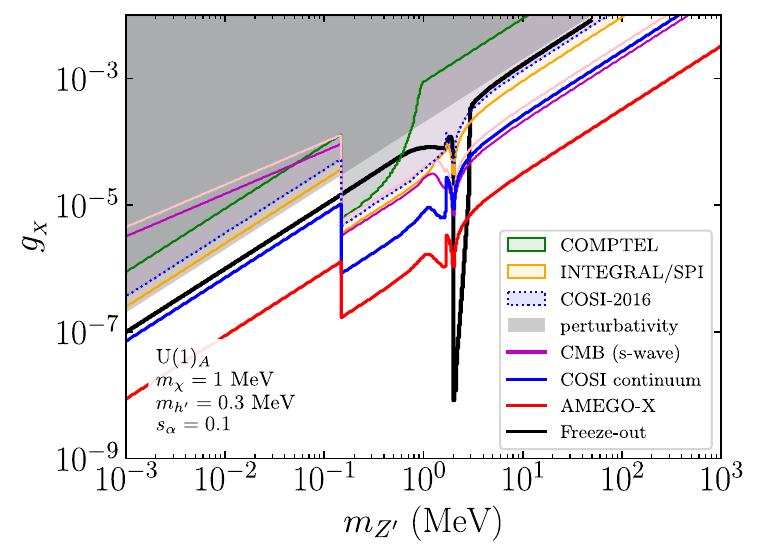}
\includegraphics[width=0.496
\textwidth]{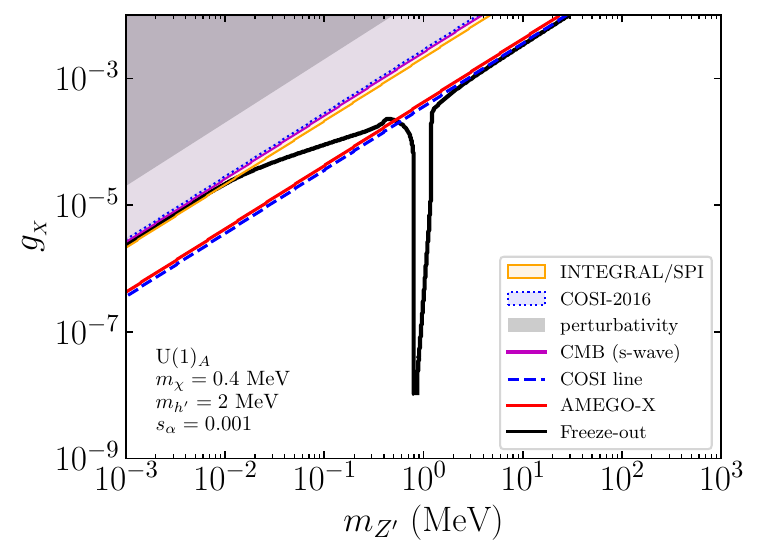} 
\includegraphics[width=0.496
\textwidth]{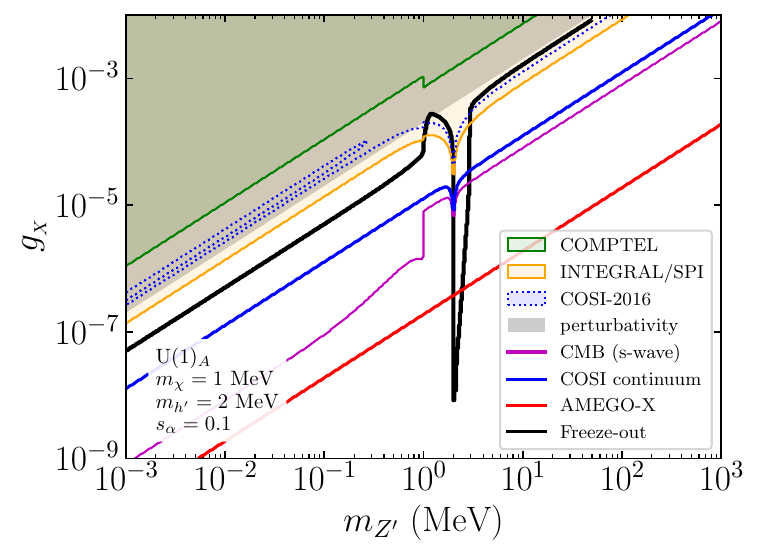}
\caption{Constraints on the $U(1)_A$ gauge coupling as function of the $Z'$ mass, for different mass hierarchies: $m_\chi > m_{h'}$ and $m_\chi < m_{h'}$ (upper and lower panels) and $m_\chi < m_e$ and $m_\chi > m_e$ (left and right panels). The indirect detection limits are color-coded as in Fig.\,\ref{fig:sv-mdm}, and we also show the most stringent theoretical limit coming from perturbative unitarity (gray region). Along the black curves, the correct relic abundance is achieved via standard freeze-out. Other mechanisms for the relic abundance requiring smaller values of $g_X$, such as freeze-in and freeze-out in non-standard cosmologies, will be probed by COSI and AMEGO-X. Note that COSI will set the first indirect detection limits on most of the viable parameter space of vector-scalar portal models, beyond the CMB s-wave bounds for $m_\chi<m_e$ and for $m_\chi > m_e > m_{Z'}, m_{h'}$. On the other hand, AMEGO-X will probe most of the remaining parameter space.}
\label{fig:mZp-gX}
\end{figure}

All the relevant processes leading to gamma rays significantly depend on the gauge coupling and mass of the new gauge boson, as well as on the different mass hierarchies. For this reason, we show in Fig.\,\ref{fig:mZp-gX} how the upper limits on the total annihilation cross-section discussed in Sec.\,\ref{sec:ID} are translated into upper limits on $g_X$, as a function of $m_{Z'}$. In the left (right) panels, we consider the case where the DM is lighter (heavier) than the electron, whereas in the upper (lower) panels, DM is heavier (lighter) than the singlet scalar. In the case of indirect detection limits, we compare the total annihilation cross-section to the current upper limits set by COMPTEL (green region), INTEGRAL/SPI (yellow region), and the 2016 COSI balloon flight (blue region), as well as to line and continuum sensitivities of the upcoming COSI telescope (dashed and solid blue curves) and the continuum sensitivity of AMEGO-X (red curve). We also show the upper limit on $g_X$ from measurements of the CMB spectra (magenta curves) for the case of s-wave dominated processes, and the regions excluded by the most stringent perturbative unitarity constraint (gray regions), corresponding to the requirement $\lambda_s < 4\pi$ (Eq.\,\ref{eq:ls}). The viable regions of our parameter space, where the correct relic abundance of our DM candidate $\chi$ is achieved via freeze-out, are shown as black curves. 

In the left panels of Fig.\,\ref{fig:mZp-gX}, we have chosen a much smaller value of the mixing angle ($s_\alpha=0.001$) compared to the right panels ($s_\alpha = 0.1$) because the perturbative unitarity limits would be otherwise stronger than the sensitivity of the future MeV gamma-ray telescopes. However, the gamma-ray signals are not affected by the value of the mixing angle in these cases. Since the gamma-ray signals have a cutoff at $E_\gamma \sim m_\chi$, we do not have limits from COMPTEL in the left panels. 

Let us first consider the \textbf{upper left panel}, where DM is slightly heavier than the singlet scalar ($m_\chi = 0.4$ MeV, $m_{h'}=0.3$ MeV). The freeze-out is dominated by $\chi\bar\chi\to h' Z'$ when $m_{Z'}\ll m_{h'}, m_\chi$, and by $\chi\bar\chi\to \nu \bar \nu$ (with equal contributions from the three flavors) when $m_{Z'}> 2 m_\chi$. When $m_{Z'} \sim 2 m_\chi = 0.8$ MeV, the DM annihilation cross-section is resonantly enhanced because of the on-shell production of $Z'$s. Thus, much smaller values of $g_X$ are needed to not overproduce DM in the early universe. The gamma-ray flux is dominated by $\chi\bar\chi\to\gamma\gamma$ and $\chi\bar\chi\to h'Z'$ as long as $m_{Z'}<2m_\chi - m_{h'} = 0.5$ MeV (when annihilation into $h'Z'$ is kinematically open). However, when $m_{Z'}>m_{h'}/2$, the cascade decay of $h'$ from the $h'Z'$ channel is strongly enhanced (see Eq.\,\ref{eq:CascadeDecay}), dominating the gamma-ray flux while allowed. For $m_{Z'}> 2m_\chi - m_{h'}$, the flux is dominated by annihilation into $h'h'$. In this regime where $m_{h'}<m_\chi<2m_e$, we have only gamma-ray lines and boxes as gamma-ray signals, and both INTEGRAL/SPI and the CMB exclude a portion of the viable parameter space when the $h'Z'$ channel is enhanced. COSI will be sensitive to the whole viable region outside the $Z'$ resonance, with enough energy resolution to distinguish the spectral features predicted by this class of models. In the case of $U(1)_{B-L}$, the region for $m_{Z'} < m_{h'}/2$ is out of the reach of the indirect detection limits, since the annihilation into $\gamma\gamma$ is negligible in the absence of axial-vector couplings.

In the \textbf{lower left panel}, the singlet scalar is heavier than the DM ($m_{h'}=2$ MeV), so the channels $h'Z'$ and $h'h'$ are not available. The gamma-ray spectra are dominated by the $\gamma\gamma$ channel while the freeze-out is dominated by $Z'Z'$ for $m_{Z'}<m_\chi$ and by $\nu \bar \nu$ otherwise. Since the $Z'Z'$ channel is highly dependent on the mixing angle, we need large values of $g_X$ for small values of $s_\alpha$ to not overproduce DM, which makes the current indirect detection bounds to exclude the viable region of this parameter space for light enough $Z'$s. The remaining part of the viable parameter space will be probed by COSI and AMEGO-X, except for the $Z'$ resonance region. Note that we do not have indirect detection limits in this parameter space in the case of $U(1)_{B-L}$. 

Let us now consider the right panels of Fig.~\ref{fig:mZp-gX}. In the upper (lower) right panel, DM is heavier (lighter) than the singlet scalar and heavier than the electron, with $m_\chi = 1$ MeV and $m_{h'}=0.3$ MeV ($m_{h'}=2$ MeV). The freeze-out is dominated by annihilation into $h'Z'$ ($Z'Z'$) and $e^+e^-$ for $m_{Z'}<2m_\chi$ and $m_{Z'}>2m_\chi$, respectively. In the \textbf{upper right panel}, the gamma-ray signals are dominated by annihilation into $e^+e^-$ and $\gamma\gamma$, except for $m_{h'}/2<m_{Z'}<2m_\chi -m_{h'}$, where the cascade decays of $h'$ from the $h'Z'$ channel dominate the gamma-ray flux. In this case, the viable region where $m_{Z'}>m_{h'}/2$ is ruled out by current indirect detection limits, except for the resonance regime. COSI and AMEGO-X will probe the remaining viable region where $m_{Z'}<m_{h'}/2$. In the \textbf{lower right panel}, the CMB bounds rule out all the viable regions outside the resonance regime. However, the FSR gamma rays coming from $\chi\bar\chi\to e^+e^-$ are so strong that AMEGO-X will probe tiny values of the new gauge coupling, entering the viable parameter space enabled by the freeze-in mechanism. Specifically, we note that for DM candidates with pure vector couplings, the correct relic abundance obtained via freeze-in takes place for $g_X\sim 10^{-9}$ and is independent of $m_{Z'}$ for $m_{Z'}\ll 2m_\chi$ \cite{Cosme:2021baj}. In such cases, AMEGO-X will start probing the freeze-in regime of DM production. The same qualitative conclusions hold for the charge assignment of $U(1)_{B-L}$ in the parameter spaces of the right panels. The only relevant difference is that the correct relic abundance is achieved for higher values of $g_X$ for $m_{Z'}<m_\chi$, rendering the viable region ruled out by the current indirect detection limits, except for the resonance regime.

The current indirect detection limits are stronger than the direct detection limits discussed in Sec.\,\ref{Sec:DDcolliders}. The stronger values of DM-electron elastic scattering cross-sections for $m_\chi \sim 1$ MeV are for $g_X \sim 10^{-4}$, which is currently allowed by direct detection (see Fig.\,\ref{fig:DD}) but already excluded by indirect detection and theoretical constraints. However, it is interesting to have in mind that for DM in the MeV-GeV range, we will be able to explore the complementarity between AMEGO-X and future direct detection experiments.


Finally, we emphasize that we explored viable regions regarding the freeze-out production assuming the standard cosmological scenario of a radiation-dominated universe during freeze-out. In the non-standard cosmological context of early matter domination \cite{Cosme:2020mck}, for instance, the correct relic abundance via freeze-out (freeze-in) is achieved for smaller (larger) couplings between DM and SM particles \cite{Dutra:2018gmv,Dutra:2021phm}, regardless of the specific DM model. Thus, in vector-scalar portal models as well as in other classes of models, we can find viable regions of a DM parameter space for smaller annihilation cross-sections. This motivates us to improve indirect detection limits on sub-GeV dark matter candidates.

\section{Comparison with previous limits}
\label{sec:literature}

Several works in the recent literature have explored light dark matter candidates, with masses in the keV-GeV range \cite{Essig:2013goa,Boddy:2015efa,Boudaud:2016mos,Gonzalez-Morales:2017jkx,Boudaud:2018oya,Laha:2020ivk,Cirelli:2020bpc,Coogan:2021rez,Coogan:2021sjs,Caputo:2022dkz,Cirelli:2023tnx,DelaTorreLuque:2023olp,Balan:2024cmq,Watanabe:2025pvc,Tang:2025vqf,Balaji:2025afr,Bernal:2025szh,Nguyen:2025tkl,Cirelli:2025rky,Wang:2025jhy}, demonstrating how much attention these candidates have gained in recent years. This growing interest is motivated by and also motivates many future MeV gamma-ray telescopes \cite{Aramaki:2022zpw}. In this work, we explored the gamma-ray signals of a class of models for fermionic light DM in which both a vector and a scalar mediators play a significant role in the DM origin and phenomenology. We have shown our results for specific realizations of such vector-scalar portal models and discussed how these results depend on the most relevant parameters, enabling the extrapolation of our findings for other model realizations. We derived the sensitivities of COSI and AMEGO-X in this work, but interested readers can modify our analysis with the {\tt Hazma} code available at {\tt GitHub} \href{https://github.com/dutramaira/VectorScalarPortal_Hazma}{\faGithub} with other DM models and telescopes. 


When DM is the lightest particle, the annihilation into $\gamma\gamma$ dominates the current indirect detection limits on the annihilation cross section posed by INTEGRAL/SPI, which overcomes the CMB limit on s-wave annihilation into $\gamma\gamma$ \cite{Laha:2020ivk}\,\footnote{Annihilation into neutrinos are much less constrained, see for instance \cite{BetancourtKamenetskaia:2025rwk}.}. We have shown that COSI line searches would probe cross-sections from ten to hundred times lower than the current limits for the first time. In Refs.\,\cite{Caputo:2022dkz,Watanabe:2025pvc}, the authors show how the COSI limits depend on the observation target and uncertainties in the DM halo profile. As we have discussed, in the context of vector-scalar portal models, line searches would be relevant only in the presence of axial-vector couplings enabling an s-wave annihilation via the effective $Z'\gamma\gamma$ interaction, since the predicted cross-sections would be otherwise negligible.


When DM is heavier than the electron, the $e^+e^-$ annihilation channel easily dominates the gamma-ray flux via FSR. In vector-scalar portals, the DM annihilation into fermions is typically s-wave because the vector exchange easily dominates (see Eq.\,\ref{eq:sv_ee}). As expected from previous results, the CMB limit on s-wave DM annihilation into $e^+e^-$ overcomes the current diffuse $\gamma$-ray limits ($\langle \sigma v \rangle \gtrsim  10^{-27}$ cm$^3$/s)~\cite{Essig:2013goa,Gonzalez-Morales:2017jkx,Cirelli:2020bpc} and COSI's sensitivity \cite{Caputo:2022dkz}. In this case, we found that COSI will only probe cross-sections below the CMB limits if strong gamma-ray boxes from annihilation into mediators are allowed.


The possibility of probing light DM particles with cosmic rays using Voyager~1 was investigated in Refs.~\cite{Boudaud:2016mos,Boudaud:2018oya,DelaTorreLuque:2023olp,Cirelli:2025rky}, leading to strong constraints on the annihilation cross section into $e^+e^-$ and other annihilation channels. However, these limits are weaker than the CMB limit on s-wave annihilation into fermions and we thus neglect them in this work. 


For DM masses above tens of MeV, the secondary photons produced from inverse Compton scattering (ICS) between energetic $e^+/e^-$s produced from DM annihilation and ambient light dominate over the prompt photon emission \cite{Cirelli:2020bpc,Cirelli:2023tnx,Balaji:2025afr,DelaTorreLuque:2023olp,Balaji:2025afr,Nguyen:2025tkl,Cirelli:2025rky}. However, the limits considering ICS are still excluded by or similar to the CMB limits on s-wave DM annihilation, so we neglect them in this work for simplicity as we focus on lower DM masses.

Future MeV gamma-ray telescopes are expected to surpass the current CMB limits on particle DM annihilation over a significant portion of the parameter space \cite{Coogan:2021rez,Coogan:2021sjs,Caputo:2022dkz,Watanabe:2025pvc,Tang:2025vqf,Cirelli:2025rky}, as we have shown in this work for the cases of COSI and AMEGO-X. From a theoretical perspective, these experiments are particularly appealing within several well-motivated extensions of the Standard Model, such as dark photon portals, right-handed neutrino dark matter, axion-like particles, and complete models involving new symmetries like $L_{\mu}-L_{\tau}$ \cite{Coogan:2021rez,Caputo:2022dkz,Balan:2024cmq,Watanabe:2025pvc,Tang:2025vqf,Nguyen:2025tkl,Cirelli:2025rky,Watanabe:2025axr,Bernal:2025szh}. They will also offer the opportunity to detect for the first time the MeV Hawking radiation from primordial black hole DM candidates, with masses around $10^{17}$g \cite{Laha:2020ivk,Coogan:2021rez,Caputo:2022dkz,Balaji:2025afr}. We thus expect to start a new era in the search for dark matter in the upcoming decade. As such, it is important to explore regions of different parameter spaces providing the correct relic abundance of dark matter that can be tested by these future experiments.

\section{Conclusions}
\label{sec:conclusions}

MeV gamma-ray astronomy will soon be revolutionized by the advent of new telescopes such as COSI, optimized for line searches with an excellent energy resolution of less than 1\% FWHM in the 200 keV to 5 MeV band, and AMEGO-X, with a continuum sensitivity in the 100 keV - 1 GeV band similar to that of Fermi-LAT. As a consequence, we will be able to probe underexplored regions of the parameter space of well-motivated dark matter (DM) models. We have studied a wide class of DM models, the vector-scalar portal models, in which the annihilation of sub-GeV fermionic DM generates gamma-ray signals at the MeV scale with distinguishable spectral features such as lines and boxes, as well as strong continuum gamma rays.  

In this work, the Standard Model (SM) fermions and a Dirac fermion DM candidate $\chi$ are charged under a local symmetry $U(1)_X$, thus interacting by exchanging the associated gauge boson $Z'$. In this widely studied simplified vector portal model, sub-GeV DM annihilate into light charged leptons and produce a continuum gamma-ray flux via final state radiation. However, in order to have gamma-ray lines in this context, both the DM and SM fermions must have non-zero axial-vector couplings to $Z'$. Interestingly, the self-consistence of this setup requires the inclusion of a singlet scalar $h'$ that breaks $U(1)_X$ and gives mass to $Z'$ and $\chi$. The mass scales of $\chi$, $Z'$, and $h'$ are thus strongly related. This class of vector-scalar portal models is much less considered in the literature but provide a rich phenomenology with gamma rays. In fact, the sub-GeV mediators enable strong gamma-ray lines (mainly from the $\gamma\gamma$ channel) and cascade annihilation leading to a box-shaped gamma-ray flux (from the $h'Z'$ and $h'h'$ channels), as shown in Fig.\,\ref{fig:spectra}. Moreover, the relic abundance via freeze-out is set by the DM annihilation into $h'Z'$ and neutrinos, with thermal cross-sections much below the canonical value and allowed by the current indirect detection limits. 

We considered the limits set by COMPTEL, INTEGRAL/SPI, and by the 46-day COSI balloon flight of 2016, done in preparation for the COSI satellite scheduled to launch in 2027. It is remarkable that the COSI balloon flight provided constraints on the DM parameter space which are comparable to the INTEGRAL/SPI limits. This shows the promising opportunities with DM searches with the COSI mission. We have also considered the CMB limits set by the Planck satellite. We estimated the discovery reach of COSI and AMEGO-X to vector-scalar portal models using as benchmarks the charge assignments of the usual $U(1)_{B-L}$ and of a pure axial $U(1)_A$. We provided general expressions for the most relevant annihilation cross-sections to help in the understanding of how the gamma-ray signals depend on the model parameters. Our results are thus easily applicable to other charge assignments and other choices of parameters. 

We discussed how the different mass hierarchies among $\chi$, $h'$, $Z'$, and the electron impact the indirect detection limits (Fig.\,\ref{fig:sv-mdm}). If DM is the lighter particle, only the $\gamma\gamma$ annihilation channel is possible and cross-sections as low as $10^{-34}$ cm$^3/$s will be probed by COSI line searches. When $m_{h'}<m_\chi<m_{Z'},m_e$, the cascade decays $h'\to\gamma\gamma$ from the $h'h'$ channel dominate the signal, and cross-sections up to $10^{-36}$ cm$^3/$s will be probed by AMEGO-X. For $m_\chi > m_e$, the COSI continuum limits can be stronger than the CMB limits for 1-10 MeV DM, but the AMEGO-X limits can be as strong as $10^{-33}$ cm$^3/$s for 1 MeV DM. Finally, when $m_\chi > m_e > m_{h'}$, the limits can be about one order of magnitude stronger because of the $h'h'$ and $h'Z'$ channels. The complementarity between indirect detection, direct detection, and colliders is crucial to validate a possible DM signal. We have shown that direct detection limits, posed by XENON1T, DarkSide-50, and SENSEI, become relevant for $m_\chi > 1$ MeV (Fig.\,\ref{fig:DD}), probing the same parameter space of vector-scalar portal models as AMEGO-X. We also plan to explore the complementarity between COSI and collider limits on generic vector-scalar portal models in a future work.

Our main results were shown in Fig.\,\ref{fig:mZp-gX}, where we show the limits on the new gauge coupling and $m_{Z'}$ for different mass hierarchies and scalar mixing. COSI will set the first indirect detection limits on most of the viable parameter space of vector-scalar portal models, beyond the CMB s-wave bounds for $m_\chi<m_e$ and for $m_\chi > m_e > m_{Z'}, m_{h'}$. While this conclusion holds for pure vector charge assignments such as in $U(1)_{B-L}$, in the case of axial-vector couplings, we can search for sub-MeV DM with gamma-ray line searches. It is worth emphasizing COSI's ability to resolve spectral features, increasing our capacity to distinguish a possible DM signal from the strong backgrounds. In this regard, the possible complementarity between COSI and planned missions with excellent angular resolution such as GECCO is very promising. We showed that the AMEGO-X limits would be strong enough to constrain even the resonance region, that usually remains viable, especially for $m_{h'} > m_\chi > m_e$ (when the annihilation branching fraction into $e^+ e^-$ is larger). The strong AMEGO-X limits might also probe the parameter space of frozen-in DM for keV-scale $Z'$ in the case of vector-like MeV-scale DM. Finally, it is important to emphasize that both COSI and AMEGO-X will probe the parameter space of DM beyond the standard freeze-out, for instance in scenarios of early matter domination, in the context of vector-scalar portals and in other classes of models.

\section*{Acknowledgments} We thank Chris Karwin, Regina Caputo, and Alexander Moiseev for helpful discussions. M.D. acknowledges support by an appointment to the NASA Postdoctoral Program at the NASA Goddard Space Flight Center,
administered by Oak Ridge Associated Universities through a contract with NASA. C.S. acknowledges support by CNPq through grants number 304944/2025-4 and 447820/2025-7, and by the S\~{a}o Paulo Research Foundation (FAPESP) through grant number 2021/01089-1.

\appendix

\section{Model implementation in {\tt Hazma}}
\label{sec:Hazma}

In this Appendix, we provide details about our implementation of a generic vector-scalar portal model in {\tt Hazma} \cite{Coogan:2019qpu}. We computed the cross-sections of the processes relevant for indirect detection (Fig.\,\ref{fig:diagrams}) in terms of the free parameters of a generic vector-scalar portal. Thus, different realizations of the $U(1)_X$ symmetry can be studied. The notebook containing the full implementation of our model, as well as the inputs for the indirect detection and CMB analyses, are available at {\tt GitHub} \href{https://github.com/dutramaira/VectorScalarPortal_Hazma}{\faGithub}.

\bigskip
To implement a new model in {\tt Hazma}, we need to provide the following functions:
\begin{itemize}
  \item {\tt annihilation\_cross\_sectio\_funcs()}: get functions that compute the cross-sections for each channel;
  \item {\tt spectrum\_funcs()}: get functions that compute the continuum gamma-ray spectrum for annihilation into each channel;
  \item {\tt gamma\_ray\_lines($E_{cm}$)}: gives the energy of the photon and the branching fraction into the channels leading to gamma-ray lines; 
  \item {\tt positron\_spectrum\_funcs()}: get functions that compute the continuum positron spectrum for annihilation into each channel;
  \item {\tt positron\_lines($E_{cm}$)}: gives the energy of the positron and the branching fraction into the channels leading to positron lines; 
\end{itemize}

\bigskip
First, we defined a class to compute the partial decay widths of $Z', h', Z$, and $h$, which are needed for the calculation of cross-sections and spectra. For instance, we have:
\begin{lstlisting}
class DecayWidths:
    def Zp_partial_widths(self,MZp, mx, mhp, gX, XXL, XPhi, XPhiS, 
                          XLL, XQL,sa)
    # call functions that compute the partial decay widths 
    # and check thresholds
        return {"x x": pwxx,
                "e e": pwee, "mu mu": pwmumu, "tau tau": pwtautau,
                "pi0 g": pwpi0g, "hp g": pwhpg,
                "u u": pwuu, "d d": pwdd, "c c": pwcc,
                "s s": pwss, "t t": pwtt, "b b": pwbb,
                "total": total }

    # Similarly for the scalar decay width:
    def hp_partial_widths(self,MZp, mx, mhp, gX, XXL, XPhi, XPhiS, 
                          XLL, XQL,sa)
\end{lstlisting}

We then defined a class containing the functions that compute the cross-sections and the required {\tt Hazma} functions {\tt annihilation\_cross\_sectio\_funcs}, {\tt annihilation\_cross\_sections}, and {\tt annihilation\_branching\_fractions}:

\begin{lstlisting}
class CrossSections:

    def sigma_XX_to_ff(self, e_cm, mf, XfL, Qf, t3f):
        mx = self.mx
        MZp = self.MZp
        mhp = self.mhp
        gX  = self.gX
        sa = self.sa
        XPhi = self.XPhi
        XPhiS = self.XPhiS
        XXL = self.XXL
        XLL = self.XLL
        XQL = self.XQL
        (...)
        # returns the cross-section for the process $\bar \chi \chi \to \bar f f$ in terms 
        # of the free parameters ($m_\chi, m_{Z'}, m_{h'}, g_X, s_\alpha, X_{\Phi}, X_{\Phi_s}, X_{\chi_L}, X_{L_L}, X_{Q_L}$)

# similarly for the other processes:
    def sigma_XX_to_pi0g(self, e_cm): # $\bar \chi \chi \to \pi^0 \gamma$
    def sigma_XX_to_gg(self,e_cm):    # $\bar \chi \chi \to \gamma \gamma$
    def sigma_XX_to_hpg(self, e_cm):  # $\bar \chi \chi \to h' \gamma$
    def sigma_XX_to_Zpg(self, e_cm):  # $\bar \chi \chi \to Z' \gamma$
    def sigma_XX_to_ZpZp(self, e_cm): # $\bar \chi \chi \to Z' Z'$
    def sigma_XX_to_hphp(self, e_cm): # $\bar \chi \chi \to h' h'$
    def sigma_XX_to_hpZp(self, e_cm): # $\bar \chi \chi \to h' Z'$  

# ----------- required Hazma functions:

    def annihilation_cross_section_funcs(self):
        return { 
        "e e": lambda e_cm: self.sigma_XX_to_ff(e_cm, me, self.XLL, 
                                                -1, -1./2),
        "mu mu": lambda e_cm: self.sigma_XX_to_ff(e_cm, mmu, self.XLL,                                    -1, -1./2),
        "pi0 g": lambda e_cm: self.sigma_XX_to_pi0g(e_cm),
        "hp g": lambda e_cm: self.sigma_XX_to_hpg(e_cm),
        "Zp g": lambda e_cm: self.sigma_XX_to_Zpg(e_cm),
        "g g": lambda e_cm: self.sigma_XX_to_gg(e_cm),
        "Zp Zp": lambda e_cm: self.sigma_XX_to_ZpZp(e_cm),
        "hp hp": lambda e_cm: self.sigma_XX_to_hphp(e_cm),
        "hp Zp": lambda e_cm: self.sigma_XX_to_hpZp(e_cm) }

    def annihilation_cross_sections(self, e_cm: float) -> Dict[str, float]:
        sigmas = {
            fs: sigma_fn(e_cm)
            for fs, sigma_fn in self.annihilation_cross_section_funcs().items()
        }
        sigmas["total"] = sum(sigmas.values())
        return sigmas

    def annihilation_branching_fractions(self, e_cm: float) -> Dict[str, float]:
        cs = self.annihilation_cross_sections(e_cm)

        if cs["total"] == 0:
            return {fs: 0.0 for fs in cs if fs != "total"}
        else:
            return {
                fs: sigma / cs["total"] for fs, sigma in cs.items() if fs != "total"
            }
\end{lstlisting}

\bigskip
The next step is to define the class containing the functions that compute the gamma-ray spectra and the required {\tt Hazma} functions {\tt spectrum\_funcs} and {\tt gamma\_ray\_lines}:

\begin{lstlisting}
  class Spectra:

# ----------- FSR
    def _dnde_xx_to_Zp_to_ffg(self, e_gams, e_cm, mf): # Eq. (3.7)
    def _dnde_xx_to_hp_to_ffg(self, e_gams, e_cm, mf): # Eq. (3.8)

# ----------- cascade FSR
    def _integrand_VectorMediator(self,x,e_cm,mi,mf):
    def _integral_VectorMediator(self,e_gam,e_cm,mi,mf):
    def _dnde_iffg_VectorMediator(self, e_gams, e_cm, mi, mf): 
    # Eq. (3.9) using Eq. (3.7)

    def _integrand_ScalarMediator(self,x,e_cm,mi,mf):
    def _integral_ScalarMediator(self,e_gam,e_cm,mi,mf):
    def _dnde_iffg_ScalarMediator(self, e_gams, e_cm, mi, mf): 
    # Eq. (3.9) using Eq. (3.8)

# ----------- cascade $\gamma \gamma$
    def _dnde_hphpgg(self, e_gam, e_cm): # Eq. (3.6) w/ $AB=h'h'$
    def _dnde_hpgg(self, e_gam, e_cm): # Eq. (3.6) w/ $AB=h'\gamma$
    def _dnde_hpZpgg_hp(self, e_gam, e_cm): # Eq. (3.6) w/ $AB=h'Z'$

# ----------- spectra
    def dnde_ee(self, e_gams, e_cm):
    # _dnde_xx_to_Zp_to_ffg + _dnde_xx_to_hp_to_ffg w/ $ff = e^+e^-$
 
    def dnde_mumu(self, e_gams, e_cm):
    # _dnde_xx_to_Zp_to_ffg + _dnde_xx_to_hp_to_ffg w/ $ff = \mu^+\mu^-$
    # + 2.0 * spectra.dnde_photon_muon(e_gams, e_cm / 2.0),
    # built-in function to compute the $\gamma$-ray spectrum from $\mu^\pm$ decays

    def dnde_hphp(self, e_gams, e_cm): # $h'h'$ channel
    # define free parameters
        widths = self.hp_partial_widths(mhp, MZp, mx, sa, gX, XXL, XPhi, XPhiS, XLL, XQL)
        return (2.0 * widths.get("g g")/widths.get("total") 
                * self._dnde_hphpgg(e_gams, e_cm) 
                + 2.0 * widths.get("e e")/widths.get("total") 
                * self._dnde_iffg_ScalarMediator(e_gams,e_cm,mhp,me) 
                + 2.0 * widths.get("mu mu")/widths.get("total") 
                * self._dnde_iffg_ScalarMediator(e_gams,e_cm,mhp,mmu))

# Similarly for the other channels involving cascade decays 
    def dnde_hpg(self, e_gams, e_cm): # $h'\gamma$
    def dnde_ZpZp(self, e_gams, e_cm): # $Z'Z'$ 
    def dnde_Zpg(self, e_gams, e_cm): # $Z'\gamma$ 
    def dnde_hpZp(self, e_gams, e_cm): # $h'Z'$
    def dnde_pi0g(self, e_gams, e_cm): # $\pi^0\gamma$
        e_pi0 = (e_cm**2 + mpi0**2) / (2.0 * e_cm)
        return spectra.dnde_photon_neutral_pion(e_gams, e_pi0)
               # built-in function to compute the $\gamma$-ray spectrum 
               # from $\pi^0$ decays

# ----------- required Hazma functions:

    def spectrum_funcs(self):
        return {"e e":   self.dnde_ee,
                "mu mu": self.dnde_mumu,
                "hp hp": self.dnde_hphp,
                "hp g":  self.dnde_hpg,
                "Zp Zp": self.dnde_ZpZp,
                "Zp g":  self.dnde_Zpg,
                "hp Zp": self.dnde_hpZp,
                "pi0 g": self.dnde_pi0g }

    def gamma_ray_lines(self, e_cm):
        bfgg   = self.annihilation_branching_fractions(e_cm)["g g"] 
        bfpi0g = self.annihilation_branching_fractions(e_cm)["pi0 g"]
        bfhpg = self.annihilation_branching_fractions(e_cm)["hp g"]
        bfZpg = self.annihilation_branching_fractions(e_cm)["Zp g"]
        return {"g g": {"energy": e_cm / 2.0, "bf": bfgg}, 
                "pi0 g": {"energy": (e_cm**2 - mpi0**2)/(2.0*e_cm),        "bf": bfpi0g},
                "hp g": {"energy": (e_cm**2 - self.mhp**2)/(2.0*e_cm),      "bf": bfhpg},  
                "Zp g": {"energy": (e_cm**2 - self.MZp**2)/(2.0*e_cm),      "bf": bfZpg} }

\end{lstlisting}

\bigskip
The {\tt Hazma} function {\tt total\_spectrum} weights the channels in {\tt spectrum\_funcs} with the annihilation branching fractions and returns the total continuum gamma-ray spectrum. The total spectrum, with continuum and line contributions, can be convolved with the energy resolution of a given telescope with the {\tt Hazma} function {\tt total\_conv\_spectrum\_fn}. 


\bigskip
We similarly define the class with the remaining required functions that compute the positron spectrum from DM annihilation, needed to derive the CMB limits:

\begin{lstlisting}
class PositronSpectra:

# ----------- cascade $e^+ e^-$
    def _dnde_pos_hpee(self, e_pos, e_cm): # Eq. (3.6) w/ $AB=h'\gamma$
    def _dnde_pos_Zpee(self, e_pos, e_cm): # Eq. (3.6) w/ $AB=Z'\gamma$
    def _dnde_hphpee(self, e_pos, e_cm): # Eq. (3.6) w/ $AB=h'h'$
    def _dnde_ZpZpee(self, e_pos, e_cm): # Eq. (3.6) w/ $AB=Z'Z'$
    def _dnde_hpZpee_hp(self, e_gam, e_cm): # Eq. (3.6) w/ $AB=h'Z'$ 
    # (scalar decay)
    def _dnde_hpZpee_Zp(self, e_gam, e_cm): # Eq. (3.6) w/ $AB=h'Z'$ 
    # (vector decay)

# ----------- cascade FSR
    def _integrand_VectorMediator(self,x,e_cm,mi,mf):
    def _integral_VectorMediator(self,e_ps,e_cm,mi,mf):
    def _dnde_pos_iffg_VectorMediator(self, e_ps, e_cm, mi, mf): 
    # Eq. (3.9) using Eq. (3.7)

    def _integrand_ScalarMediator(self,x,e_cm,mi,mf):
    def _integral_ScalarMediator(self,e_ps,e_cm,mi,mf):
    def _dnde_iffg_ScalarMediator(self, e_ps, e_cm, mi, mf): 
    # Eq. (3.9) using Eq. (3.8)

# ----------- spectra 
    def dnde_pos_mumu(self, e_ps, e_cm):
        return spectra.dnde_positron_muon(e_ps, e_cm/2.0)
        # built-in function to compute the positron spectrum 
        # from $\mu^\pm$ decays

    # cascade $e^+ e^-$ + cascade FSR from $h'$ and/or $Z'$ decays:
    def dnde_pos_hpg(self, e_ps, e_cm): # $h'\gamma$ 
    def dnde_pos_Zpg(self, e_ps, e_cm): # $Z'\gamma$ 
    def dnde_hphp(self, e_ps, e_cm): # $h'h'$ 
    def dnde_ZpZp(self, e_gams, e_cm): # $Z'Z'$ 
    def dnde_hpZp(self, e_gams, e_cm): # $h'Z'$

# ----------- required Hazma functions:

    def positron_spectrum_funcs(self):
        return {"mu mu": self.dnde_pos_mumu,
                "hp g":  self.dnde_pos_hpg,
                "Zp g":  self.dnde_pos_Zpg,
                "hp hp": self.dnde_pos_hphp,
                "Zp Zp": self.dnde_pos_ZpZp,
                "hp Zp":  self.dnde_pos_hpZp }

    def positron_lines(self, e_cm):
        bf = self.annihilation_branching_fractions(e_cm)["e e"] 
        return {"e e": {"energy": e_cm/2.0, "bf": bf} }

\end{lstlisting}

\bigskip
The implementation of new direct and cascade annihilation final states that could contribute to the gamma-ray and positron spectra is relatively straightforward. A potential limitation of our current implementation is the absence of hadronic direct and cascade annihilation channels relevant for DM masses above hundreds of MeV, as it may impact the sensitivity of future pair-telescopes on the parameter space.     

\bigskip
Having defined the classes for computing decay widths, cross-sections, and spectra, we can define the model class inheriting all the functions from the {\tt Hazma} classes {\tt TheoryAnn} and {\tt TheoryCMB}. With that, we can convolve the spectra ({\tt total\_conv\_spectrum\_fn}) and derive the limits from indirect detection searches ({\tt binned\_limit}) and the CMB limits ({\tt cmb\_limit}). The model class is defined as follows:

\begin{lstlisting}
class VectorScalarPortal(DecayWidths, CrossSections, Spectra, PositronSpectra, TheoryAnn, TheoryCMB): 
        
    def __init__(self, mx, MZp, mhp, gX, sa, XPhi, XPhiS, XXL,      
                 XLL, XQL, Lam):
        self.mx = mx
        self.MZp = MZp
        self.mhp = mhp
        self.gX = gX
        self.sa = sa
        self.XPhi = XPhi
        self.XPhiS = XPhiS
        self.XXL = XXL
        self.XLL = XLL
        self.XQL = XQL
        self.Lam = Lam

    @staticmethod
    def list_annihilation_final_states():
        return ["e e", "mu mu", "pi0 g", "hp g", "Zp g", "g g", 
                "Zp Zp", "hp hp", "hp Zp"]
\end{lstlisting}  

\bigskip
The model can be instantiated with
\begin{lstlisting}
params = {'mx':10.0, 'MZp': 8.0, 'mhp': 1.0, 'gX': 1e-4, 'sa': 1e-2, 
'XPhi': 0, 'XPhiS': 2, 'XXL': 3, 'XLL': -1, 'XQL': 1/3, 'Lam': 1e6}

model = VectorScalarPortal(**params)
\end{lstlisting} 

\bigskip
This implementation allows to use all methods built in {\tt Hazma}. For instance, we can:
\begin{itemize}
  \item set the free parameters of the model, and choose the $U(1)_X$ realization: \\ model.mx, model.XPhi, etc;
  \item calculate the cross-section of a given channel or the total cross-section: \\ 
  e.g., {\tt model.annihilation\_cross\_sections(e\_cm)["e e"]} \\
  and {\tt model.annihilation\_cross\_sections(e\_cm)["total"]};
  \item calculate the total convolved gamma-ray or positron spectrum, \\
  e.g., {\tt model.total\_conv\_spectrum\_fn(e\_gam\_min, e\_gam\_max, e\_cm, \\ energy\_res=energy\_res\_func,n\_pts=1000)};
  \item calculate the upper limit on $\langle \sigma v \rangle$ from a gamma-ray telescope: \\ e.g., {\tt model.binned\_limit(cosi2016\_diffuse\_einasto)}; 
  \item calculate the upper limit on $\langle \sigma v \rangle$ from CMB observations: \\
  e.g., {\tt model.cmb\_limit(x\_kd)}.
\end{itemize}  

Let us now compare our vector-scalar portal to the simplified scalar and vector portals built-in models of {\tt Hazma} \cite{Coogan:2019qpu}. They have also considered a Dirac fermion DM and a singlet scalar, but the vector mediator has only pure vector couplings. 

\begin{figure}[t!]
    \centering
\includegraphics[width=0.496\textwidth]{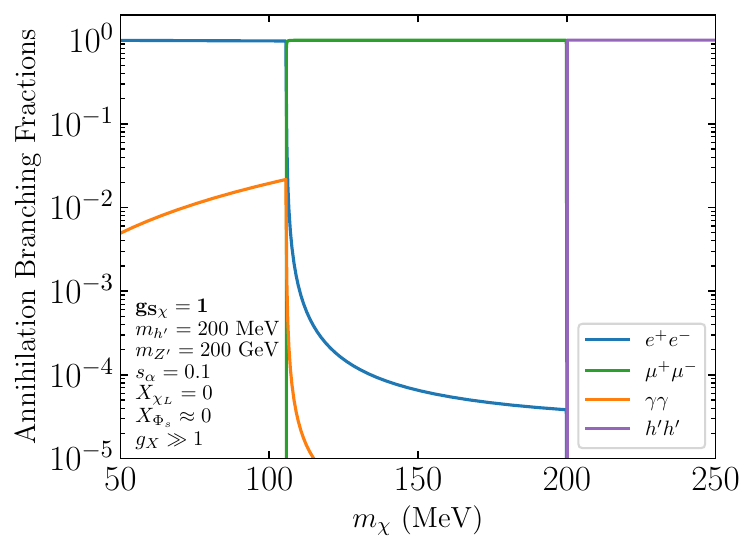} 
\includegraphics[width=0.496\textwidth]{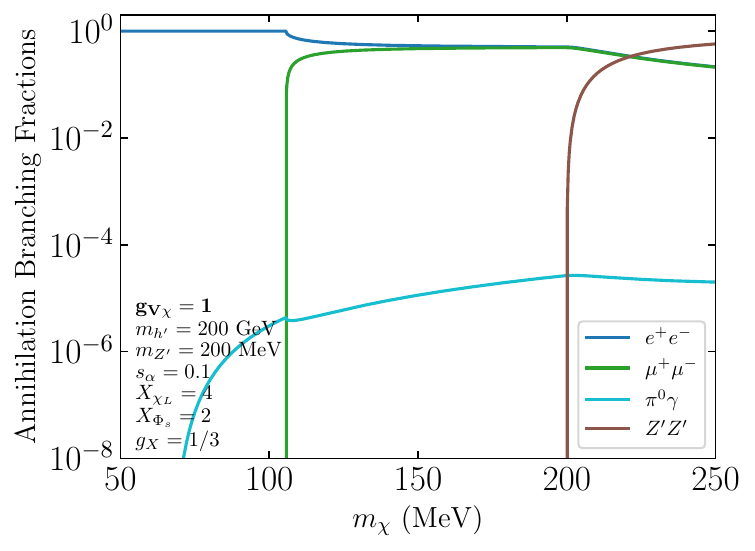} 
\caption{Comparison between our vector-scalar portal with the pure scalar (left) and vector (right) simplified portals provided in the {\tt Hazma} package \cite{Coogan:2019qpu}.}
\label{fig:simplified}
\end{figure}

In Fig.\,\ref{fig:simplified}, we show the DM annihilation branching fractions in the limiting cases of a pure scalar portal (left) and a pure vector portal (right). To reproduce the parameter choices of Figs. 3 and 4 of Ref.\,\cite{Coogan:2019qpu} (top left panels), we imposed the constraints $g_{S\chi} = 1$ and $g_{V\chi} = 1$ for the DM overal coupling to the scalar and vector, respectively. We set the corresponding mediator mass to $200$ MeV and the second mediator mass to $200$ GeV. We set $X_\Phi = 0$, $X_{L_L} = -1$, and $X_{Q_L} = 1/3$ in both cases. In our model, we assume a Yukawa-like coupling between DM and the singlet scalar giving mass to $Z'$, $g_{S\chi} = c_\alpha m_\chi/v_s$, so the constraint $g_{S\chi} = 1$ is equivalent to $g_X = m_{Z'}/(c_\alpha X_{\Phi_s} m_\chi)$. Thus, the limit case of a simplified scalar mediator, in which the vector mediator is taken to be very heavy, has no physical meaning in our framework as it would require too large values for the gauge coupling. However, we can consider DM and the singlet scalar neutral under $U(1)_X$ ($X_{\chi_L} =0$, $X_{\Phi_s} \approx 0$ \footnote{We cannot set $X_{\Phi_s} = 0$ without changing our implementation explicitly because $v_s \propto 1/X_{\Phi_s}$.}) and assume $g_{S\chi}$ as a free parameter to compare our implementation with the {\tt Hazma} scalar portal. Regarding the vector portal, imposing $g_{V\chi} = c_\xi g_X V_\chi = 1$ for our benchmark charges for $U(1)_{B-L}$ is equivalent to $g_X = 1/3$. Apart from the absence of channels with two pions, which we chose to neglect as discussed, we recover the simplified portal models. In both cases, if one allows the second mediator to be produced only by lowering their masses, the channel $h'Z'$ dominates over $h'h'$ and $Z'Z'$. 

\printbibliography

\end{document}